\documentclass[letterpaper,12pt]{article}
\usepackage{tabularx} % extra features for tabular environment
\usepackage{amsmath}  % improve math presentation
\usepackage{amsthm}
\usepackage{bm}
\usepackage{amssymb}
\usepackage{amsfonts}
\usepackage{graphicx}
\usepackage{subcaption}  % サブキャプションのためのパッケージ
\usepackage{xcolor}
\usepackage[margin=1in,letterpaper]{geometry} % decreases margins
\usepackage{cite}
\usepackage{hyperref}
\usepackage{enumerate}
\theoremstyle{plain}
\theoremstyle{definition}

\newcommand{\mrm}[1]{\mathrm{#1}}
\newcommand{\ket}[1]{\vert #1 \rangle}

\newcommand{\braket}[1]{\langle #1 \rangle}

\newcommand{\eq}[1]{\begin{equation}\begin{split}#1 \end{split}\end{equation}}

\newcommand{\eqs}[1]{\begin{align} #1 \end{align}}
\newcommand{\eqsnn}[1]{\begin{align*} #1 \end{align*}}
\newcommand{\order}[1]{\mathcal{O}\left( #1\right) }

\begin{document}

\title{Exact density profile in a tight-binding chain with dephasing noise}
\author{Taiki Ishiyama\footnote{ishiyama.t.d861@m.isct.ac.jp}, Kazuya Fujimoto\footnote{fujimoto.k.4927@m.isct.ac.jp}, 
Tomohiro Sasamoto\footnote{sasamoto@phys.sci.isct.ac.jp} \vspace{2mm} \\
Department of Physics, Institute of Science Tokyo,\\ 2-12-1 Ookayama, Meguro-ku, Tokyo 152-8551, Japan}
\date{\today}
\maketitle
\begin{abstract}
We theoretically investigate the many-body dynamics of a tight-binding chain with dephasing noise on the infinite interval.
We obtain the exact solution of an average particle-density profile
for the domain wall and the alternating initial conditions via the Bethe ansatz,
analytically deriving the asymptotic expressions for the long time dynamics.
For the domain wall initial condition, we obtain the scaling form of the average density, 
elucidating that the diffusive transport always emerges in the long time dynamics if the strength of the dephasing, no matter how small, is positive.
For the alternating initial condition, 
our exact solution leads to the fact that the average density displays oscillatory decay 
or over-damped decay depending on the strength of the dissipation.
Furthermore, we demonstrate that the asymptotic forms approach those of the symmetric simple exclusion process,
identifying corrections from it.

\end{abstract}

%---------------------------------------------------------------------------------------------------------------
\section{Introduction}
Recently open quantum many-body systems have been attracting increasing interests, showing rich variety of non-equilibrium phenomena, such as non-Hermitian skin effect\cite{Lin_topological,Zhang_review}, measurement-induced phase transitions~\cite{Cao_entanglement,Alberton_entanglement,Fava_nonlinear,Poboiko_theory}, and time crystals\cite{Sacha_time,Else_discrete}, which still await better theoretical understanding. Currently, such open systems have become experimentally accessible 
thanks to state-of-the-art experimental techniques in ultracold atoms and trapped ions. Hence, it is of importance to uncover intriguing physics unique to open quantum many-body systems from both theoretical and experimental perspectives.  

One of the theoretical frameworks for open quantum many-body systems is the Gorini-Kossakowski-Sudarshan-Lindblad (GKSL) equation
\cite{Lindblad, Gorini, Breuer}, which provides a legitimate description of systems when their time evolution is approximately Markovian. 
In spite of the simplification, the GKSL equation captures a variety of interesting phenomena in open quantum many-body systems and 
its theoretical predictions have been compared with several experiments
\cite{Luschen_signatures,Vatre_dynamics,Sponselee_dynamics,Honda_observation}.

It is in general difficult to solve the GKSL equation for many-body systems.
However, several exact results have been obtained, offering fundamental insights into the properties of open quantum many-body systems.
A typical example for exact solutions is non-equilibrium steady states for boundary-driven systems, for which previous literature has constructed the exact density matrix \cite{Znidaric_matrix,Znidaric_exact, Znidaric_solvable, Karevski_quantum,Prosen_exact, Prosen_open, 
Prosen_exact_hubbard,Prosen_comments,Popkov_infinitely,Popkov_exact,Popkov_solution, 
Ilievski_dissipation,Karevski_exact,
Popkov_inhomogeneous,Matsui_exact,Popkov_bethe,Lingenfelter_exact} 
and has calculated their correlations \cite{Znidaric_large,Znidaric_exact_large,Buca_connected}. 
As another example, there is a certain class of models which admits a reduction to free fermions or free bosons, for which the spectra of the time-evolution generator in the GKSL equation are calculated \cite{Prosen_third,Prosen_quantization,Prosen_spectral,Kitahama_jordan,Zheng_exact,
Guo_solutions,Guo_analytical,Barthel_solving,Shackleton_exactly,Prosen_exact_xy,Horstmann_noise,Shibata_kitaev,Yang_liouvillian}
and even the dynamics of physical quantities have been investigated 
\cite{Prosen_exact_xy,Horstmann_noise,Clark_exact,Yamanaka_exact,Shibata_kitaev,Krapivsky_free_fermion,Krapivsky_free_boson,
Alba_noninteracting,Alba_spreading,Keck_dissipation,Maity_growth,Teretenkov_exact,Vernier_mixing,Yang_liouvillian,Eisert_noise}. 
However, studying dynamical properties of physical quantities beyond such free systems is still quite challenging, 
although the spectra for certain models can be exactly obtained using the Bethe ansatz
\cite{Medvedyeva_exact, Rubio_integrability, Claeys_dissipative, Ziolkowska_yang, De_constructing, Essler_integrability, 
Nakagawa_exact, Buca_XXZ, Rrobertson_exact, Ekman_liouvillian,Shibata_ashkin,Yamamoto_universal,Yamamoto_universal_SU(N)}.

The Bethe ansatz is a powerful method for studying a certain class of one-dimensional many-body systems that cannot be reduced to free systems 
\cite{Korepin_quantum,Gaudin_bethe,Takahashi_thermodynamics}.
Many equilibrium and non-equilibrium properties have been understood by the method. 
While it is relatively easy to get information about the spectrum of the Bethe ansatz solvable systems, 
which represent energy eigenvalues of quantum Hamiltonians or 
relaxation properties of the GKSL equation, studying time dependence of physical quantities, whether they are closed or open, are much demanding 
because they would require summing up all eigenstates,
which is in many cases intractable. For time-dependent quantities 
of open quantum systems, there have been almost no exact results, except for some studies for few-body systems \cite{Eisler_ballistic,Esposito_exactly,Esposito_emergence,Marche_universality,Alba_free}. 

In the field of stochastic interacting systems, there has been an interesting development, in which several important time-dependent quantities
have been calculated exactly using the Bethe ansatz. For example for the asymmetric simple exclusion process (ASEP), which is a paradigmatic model in the area, 
full counting current distribution has been calculated by summing the Green's function constructed by the Bethe ansatz
\cite{Tracy_asymptotics,Borodin_duality}. 
A novel feature of this approach is that,
rather than studying a finite system and then taking the infinite system limit,  one uses the Bethe ansatz directly on the infinite lattice. 
Then, it is an interesting question to ask if a similar approach can be applied to open quantum many-body systems described by the GKSL equation. 

In this paper  we will show that it is the case by investigating a tight-binding chain with dephasing noise on the infinite interval, which is known to be
mappable to the one-dimensional Fermi-Hubbard model with imaginary interaction strength\cite{Medvedyeva_exact} 
and is thus integrable by the nested Bethe ansatz\cite{Lieb_hubbard,Essler_hubbard}.
We focus on the many-body dynamics of a two-point correlation function, 
particularly its diagonal element, i.e., the average particle density.
Employing the Bethe ansatz on the infinite interval, we first obtain the Green's function for the two-point correlation function,
which may be used for studying a wide class of initial conditions. 
In this paper we focus on two many-body initial conditions, namely, the domain wall and the alternating initial conditions,
and exactly calculate the average particle density by taking appropriate sums of the Green's function.  
In both cases, we determine its asymptotic form in the long time limit.
For the domain wall initial condition, we show that the asymptotic form obeys the diffusive scaling.
For the alternating initial condition, we analytically demonstrate the dynamical transition in the relaxation of 
the average particle density which was numerically found in Ref.~\cite{Haga_quasi}.

The leading terms of the asymptotic behaviors of the density agree with those in the symmetric simple exclusion process (SEP),
a well-known classical Markov process. Though connections of our model to SEP have been discussed in literature \cite{Znidaric_large,Carollo_fluctuating,Cai_algebraic},
our exact formulas give its firm confirmation, in particular for the case of the finite dephasing strength, 
and also allow us to identify the corrections from SEP. 
We remark that the appearance of SEP is not related to the fact that we employ the techniques developed in the field of stochastic processes. 

The rest of the paper is organized as follows. 
In Sec.~\ref{sec:setup}, we define the model and review the result\cite{Medvedyeva_exact} about the integrability of the model.
In Sec.~\ref{sec:integral}, we derive the Green's function for the two-point correlation function.
In Sec.~\ref{sec:exact},
we derive the exact solution of the average particle density
for the domain wall and the alternating initial conditions using the Green's function, 
obtaining its asymptotic form in the long time limit.
In Sec.~\ref{sec:correction}, we compare the derived asymptotic forms 
with those of SEP to identify corrections.

%---------------------------------------------------------------------------------------------------------------
\section{Setup and method}\label{sec:setup}
\begin{figure}[htbp]
    \centering
    \includegraphics[width=0.9\textwidth]{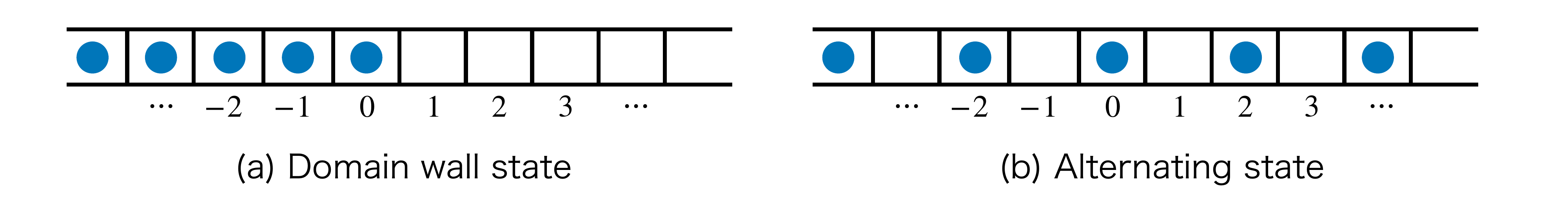} 
    \caption{Schematic illustrations of the initial states.
    The blue circles represent fermions and the numbers below the lattice denote the site numbers. 
    (a) Domain wall state $\vert \mathrm{DW}\rangle := \prod_{j\leq 0}a^\dagger_j \vert 0\rangle$. 
    The fermions occupy the left half of the system.
    (b) Alternating state $\vert \mathrm{ALT}\rangle := \prod_{j\in \mathbb{Z}} a^\dagger_{2j} \vert 0\rangle$.
    The fermions occupy every other site.}
    \label{fig:initial_dw_alt} 
\end{figure}

We consider a tight-binding chain with dephasing noise on the infinite interval.
Under the Markov approximation, the time evolution of the density matrix $\rho(t)$ 
is given by the GKSL equation\cite{Lindblad, Gorini, Breuer},
\begin{equation}
\frac{d \rho(t)}{d t} = \mathcal{L}\rho(t)
:=-i [H,\rho(t)] +
 \sum_{x \in \mathbb{Z}}\left[ L_x \rho(t) L^\dagger_{x} -\frac{1}{2} \{ L^\dagger_x L_x, \rho(t) \}\right], \label{eq:GKSL}
\end{equation}
where the superoperator $\mathcal{L}$ is referred to as the Liouvillian.
The system Hamiltonian is given by the one-dimensional tight-binding model,
$
H := -\sum_{x\in \mathbb{Z}} (a^\dagger_x a_{x+1} +a^\dagger_{x+1}a_x),
$
where $a_x (a^\dagger_x)$ is the annihilation (creation) operator of a fermion at site $x$.
The Lindblad operator $L_x := \sqrt{4\gamma}n_x$ describes dephasing caused by
the coupling to environments\cite{Breuer}.
Here, $\gamma$ and $\hat{n}_x := \hat{a}^{\dagger}_x \hat{a}_x$ represent the strength of the dephasing and the particle number operator at site $x$, respectively. 
As initial states, we will consider the domain wall state $\vert \mathrm{DW}\rangle := \prod_{j\leq 0}a^\dagger_j \vert 0\rangle$
and the alternating state $\vert \mathrm{ALT}\rangle := \prod_{j\in \mathbb{Z}} a^\dagger_{2j} \vert 0\rangle$
(see Fig.~\ref{fig:initial_dw_alt}).
The quantity of our main interest is an average particle density $\langle n_x \rangle_t:=\mathrm{Tr}[n_x \rho(t)]$, 
which is given by the diagonal element of a two-point correlation function 
$G_{x_1,x_2}(t):= \mathrm{Tr}[a^\dagger_{x_1}a_{x_2}\rho(t)]$.
The equation of motion for $G_{x_1,x_2}(t)$ reads 
\begin{equation}
i\frac{d}{dt} G_{x_1,x_2} = G_{x_1+1,x_2} + G_{x_1-1,x_2} -G_{x_1,x_2+1} -G_{x_1,x_2-1}
+4i\gamma(\delta_{x_1, x_{2}} -1) G_{x_1,x_2}, \label{eq:evo_2point}
\end{equation}
which can be derived by Eq.~\eqref{eq:GKSL}
\cite{Medvedyeva_exact,Eisler_ballistic,Zunkovic_closed,Mesterhazy_solvable,Fujimoto_impact,Barthel_solving,Nigro_competing}.
Here, we emphasize that the differential equation is closed and thus one can obtain the two-point correlation 
function without information on higher-order correlation functions. 
General criteria for such closed hierarchy for the correlation functions have been discussed in 
Ref.~\cite{Zunkovic_closed}. If we can solve the differential equation of Eq.~\eqref{eq:evo_2point}, 
the time evolution of $\langle{n_x}\rangle_t$, 
the quantity of our primary interest, is completely determined analytically. 

%-------------------------------------------------------------------------------------
%------------------------------------------------------------------------------------

%---------------------------------------------------------------------------------------------------------------
Indeed, as shown in Ref.~\cite{Medvedyeva_exact}, Eq.~\eqref{eq:evo_2point} can be mapped to 
the Schr\"{o}dinger equation for the Fermi-Hubbard model with imaginary interaction, which is Bethe ansatz solvable.
If we set
\begin{equation}
\psi_t(x_1,x_2;a_1,a_2):= 
\begin{cases}
(-1)^{x_1}G_{x_1,x_2}(t)   &\text{for}\;\; (a_1,a_2)= (\downarrow,\uparrow),\\
-(-1)^{x_2}G_{x_2,x_1}(t) &\text{for}\;\; (a_1,a_2) =(\uparrow,\downarrow),
\end{cases}\label{eq:map_to_wave}
\end{equation}
Eq.~\eqref{eq:evo_2point} is written as 
\begin{equation}
i\frac{d}{dt}\psi_t (x_1,x_2;a_1,a_2) = H_2 \psi_t (x_1,x_2;a_1,a_2),
\label{eq:hubbard}
\end{equation}
where 
\begin{equation}
H_2 := -\sum_{j=1,2}( \Delta^{+}_{j}+\Delta^{-}_{j})+4i\gamma(\delta_{x_1,x_2}-1)\label{eq:hubbard_hamiltonian}
\end{equation}
with the shift operator $\Delta^{\pm}_{j}\psi_{t}(x_1,x_2;a_1,a_2):= \psi_{t}(x_1\pm \delta_{j,1},x_2\pm\delta_{j,2};a_1,a_2)$.
One observes that $H_2$ is nothing but the two-particle Hubbard Hamiltonian with pure-imaginary interaction strength
 in first quantization (see, for example, Sec.~3.1 in Ref.~\cite{Essler_hubbard}).
In addition  $\psi_{t}(x_1,x_2;a_1,a_2)$ is antisymmetric with respect to simultaneous exchange of spin and space coordinates  
by the definition in Eq.~\eqref{eq:map_to_wave}
and can be regarded as a wave function for two fermions.
Thus, the problem of calculating the two-point correlation function for the GKSL equation with the dephasing
is reduced to solving the two-particle Schr\"{o}dinger equation for the Fermi-Hubbard model.

%-------------------------------------------------------------------------------------------------------------------------
The one-dimensional Fermi-Hubbard model is known to be
exactly solvable by the nested Bethe ansatz\cite{Lieb_hubbard,Essler_hubbard}. 
Although the interaction strength is purely imaginary in Eq.~\eqref{eq:hubbard_hamiltonian}, 
the integrability of the Fermi-Hubbard model is not spoiled
\cite{Lieb_hubbard,Essler_hubbard,Medvedyeva_exact}.
The singlet solution of the stationary Schr\"{o}dinger equation $H_2\phi(x_1,x_2;a_1,a_2)=E\phi(x_1,x_2;a_1,a_2)$
is given by the following Bethe wave function,
\begin{equation}
\phi(x_1,x_2;a_1,a_2|z_1,z_2) := \phi_{a_1,a_2}\times(z_1^{x_{Q(1)} }z_2^{x_{Q(2)} } + \frac{s_1-s_2+2\gamma}{s_1-s_2-2\gamma} z_2^{x_{Q(1)}} z_1^{x_{Q(2)}})
\label{eq:vector_bethe} 
\end{equation}
with two  complex variables $z_1,z_2$. Here we use the notations 
$\phi_{a_1,a_2}:= (\delta_{a_1,\downarrow}\delta_{a_2,\uparrow}-\delta_{a_1,\uparrow}\delta_{a_2,\downarrow})/2 $,  
$E(z_1,z_2):= -2\sum_{j=1,2}c_j -4i\gamma$, 
$s_j := (z_j -1/z_j)/2i,\; c_j := (z_j+1/z_j)/2$, and the permutation $Q$ such that $x_{Q(1)}\leq x_{Q(2)}$.
We refer the reader to Sec.~3.2 in Ref.~\cite{Essler_hubbard} for the derivation of this expression.
On the finite lattice, the boundary condition gives the Bethe equations which determine possible values of $z_1,z_2$. 
In this paper we consider the infinite lattice and do not use them. Instead we utilize 
Eq.~\eqref{eq:vector_bethe} to find a contour integral formula 
for the Green's function. 

%---------------------------------------------------------------------------------------------------------------
\section{Green's function for the two-point correlation function}\label{sec:integral}
In this paper the Green's function $\mathcal{G}_{x_1,x_2}(t;y_1,y_2)$ for Eq.~\eqref{eq:evo_2point} is 
defined as the solution of Eq.~\eqref{eq:evo_2point} for an initial condition
$\mathcal{G}_{x_1,x_2}(0;y_1,y_2)= \delta_{x_1,y_1}\delta_{x_2,y_2}$.
Once this is obtained, the two-point correlation function $G_{x_1,x_2}(t)$ for any initial density matrix $\rho_0$ can be expressed
as $G_{x_1,x_2}(t)=\sum_{y_1,y_2} \mathcal{G}_{x_1,x_2}(t;y_1,y_2) G_{y_1,y_2}(0)
=\sum_{y_1,y_2} \mathcal{G}_{x_1,x_2}(t;y_1,y_2) \mathrm{Tr}[a^\dagger_{y_1}a_{y_2}\rho_0]$.
For the two initial states considered in this work, namely the domain wall and the alternating initial states, 
one has $G^{\mathrm{DW}}_{x_1,x_2}(0) = \sum_{y\leq 0} \delta_{x_1,y}\delta_{x_2,y}$ and
$G^{\mathrm{Alt}}_{x_1,x_2}(0) = \sum_{y\in \mathbb{Z}} \delta_{x_1,2y}\delta_{x_2,2y}$. 
Hence the two-point correlation function can be expressed as follows,
\begin{equation}
G^{\mathrm{DW}}_{x_1,x_2}(t) = \sum_{y\leq 0} \mathcal{G}_{x_1,x_2}(t;y,y), ~~
G^{\mathrm{Alt}}_{x_1,x_2}(t) = \sum_{y\in \mathbb{Z}}\mathcal{G}_{x_1,x_2}(t; 2y, 2y).\label{eq:expansion_alt}
\end{equation}
Thus, in the following, we focus on $\mathcal{G}_{x_1,x_2}(t;y,y)$. 
The same quantity on a finite lattice for a slightly different dissipation was analyzed in Ref.~\cite{Eisler_ballistic}. 

Let $\psi_{t}(x_1,x_2;a_1,a_2;y)$ denote the solution of the Schr\"{o}dinger equation~\eqref{eq:hubbard} for the initial condition,
\begin{equation}
\psi_t(x_1,x_2;a_1,a_2;y)\vert_{t=0} 
= (\delta_{a_1,\downarrow}\delta_{a_2,\uparrow}-\delta_{a_1,\uparrow}\delta_{a_2,\downarrow})\delta_{x_1,y}\delta_{x_2,y} 
\label{eq:initial_hubbard}
\end{equation}
corresponding to the  doubly occupied state at the site $y$.
By Eq.~\eqref{eq:map_to_wave}, we have 
\begin{equation}
\mathcal{G}_{x_1,x_2}(t;y,y) = (-1)^{x_1-y}\psi_{t}(x_1,x_2;\downarrow,\uparrow;y).\label{eq:green}
\end{equation}

%-------------------------------------------------------------------------------------------------------------------
Now a key observation in this paper is that this quantity admits an exact and explicit expression in the form of a double contour integral,
\begin{equation}
\psi_{t}(x_1,x_2;a_1,a_2;y)= 
\oint_{|z_1|=r} \frac{dz_1}{2\pi iz_1} \oint_{|z_2|=1}\frac{dz_2}{2\pi i z_2}  (z_1z_2)^{-y}e^{-iE(z_1,z_2)t}
\phi(x_1,x_2;a_1,a_2|z_1,z_2)\label{eq:integral}.
\end{equation}
Here $\phi$ and $E$ are given in and below Eq.~\eqref{eq:vector_bethe} and
the radius of the $z_1$-contour is taken to be sufficiently small, $r\ll1$,
so that the poles of $1/(s_1-s_2-2\gamma)$ in $\phi(x_1,x_2;a_1,a_2|z_1,z_2)$ are outside the $z_1$-contour.

Note that the integral formula, Eq.~\eqref{eq:integral} is valid for any $\gamma\in\mathbb{C}$ although the case $\gamma \geq 0$ is relevant to our model.

\medskip
\noindent
\textbf{Proof of Eq.~\eqref{eq:integral}.}~
Our proof consists of the following two steps;
\begin{enumerate}[(a)]
\item RHS of Eq.~\eqref{eq:integral} satisfies Eq.~\eqref{eq:hubbard},
\item RHS of Eq.~\eqref{eq:integral} satisfies the initial condition, Eq.~\eqref{eq:initial_hubbard}.
\end{enumerate}
\textbf{Step (a)}. This can be proved by noting the fact that $e^{-iE(z_1,z_2)t}\phi(x_1,x_2;a_1,a_2|z_1,z_2)$ is 
the solution of Eq.~\eqref{eq:hubbard} from the nested Bethe ansatz\cite{Lieb_hubbard,Essler_hubbard}.
\vspace{1em}

\noindent
\textbf{Step (b)}. 
Substituting Eq.~\eqref{eq:vector_bethe} into Eq.~\eqref{eq:integral} yields
\begin{equation}
\begin{split}
\psi_{t}(x_1,x_2;a_1,a_2;y)\vert_{t=0}
=\phi_{a_1,a_2}\times
\oint_{|z_1|=r} \frac{dz_1}{2\pi iz_1} \oint_{|z_2|=1}\frac{dz_2}{2\pi i z_2} 
\left[z_1^{x_{Q(1)}-y}z_2^{x_{Q(2)}-y} + z_2^{x_{Q(1)}-y}z_1^{x_{Q(2)}-y}\right]
\\
+\phi_{a_1,a_2}\times \oint_{|z_1|=r} \frac{dz_1}{2\pi iz_1} \oint_{|z_2|=1}\frac{dz_2}{2\pi i z_2} 
\frac{4\gamma}{s_1-s_2-2\gamma}z_2^{x_{Q(1)}-y} z_1^{x_{Q(2)}-y}.
\end{split}\label{eq:proof_b}
\end{equation} 
The first term in Eq.~\eqref{eq:proof_b} gives the initial condition.
Hence, one needs to show the second term is~$0$. 
Making the substitution $z_2\to z_2/z_1$, we have
\begin{equation*}
\text{the second term in Eq.~\eqref{eq:proof_b}} = \phi_{a_1,a_2}\times \oint_{|z_1|=r} \frac{dz_1}{2\pi i} \oint_{|z_2|=r}\frac{dz_2}{2\pi i z_2}
\frac{8i\gamma\; z_1^{x_{Q(2)}-x_{Q(1)}} z_{2}^{x_{Q(1)}-y}}{-1+z_1^2-z_2+z_1^2/z_2-4i\gamma z_1}. 
\end{equation*}
Fixing $z_2$, we first integrate with respect to $z_1$.
Since we have $|z_2|=r$ on the $z_2$-contour, it follows that
$z_1^2 - z_2 + z_1^2/z_2-4i\gamma z_1=\mathcal{O}(r)$ inside the $z_1$-contour.
This implies the term $1/(-1+z_1^2-z_2+z_1^2/z_2-4i\gamma z_1)$ is holomorphic inside the $z_1$-contour. 
The term $z_1^{x_{Q(2)}-x_{Q(1)}}$ is also holomorphic since $x_{Q(1)}\leq x_{Q(2)}$.
Therefore, upon integrating with respect to $z_1$, one can conclude that the second term is $0$  by the residue theorem.
This establishes (b) and finishes the proof of Eq.~\eqref{eq:integral}.

\medskip
As demonstrated in the next section, 
the integral formula of Eq.~\eqref{eq:integral} 
enables us to derive a formula for the full time dependence of $\langle n_x\rangle_t$ 
and study its long time behaviors under the many-body initial conditions.
Note that on a finite lattice it would be difficult to find a formula like Eq.~\eqref{eq:integral} because one has to solve the Bethe equations. 
Here we treat the infinite system from the outset, 
and thus the Bethe equations are not necessary and contributions from all eigenstates are taken into account by 
taking the appropriate contours. 

%----------------------------------------------------------
\section{Exact solutions of average density profile}\label{sec:exact}
In this section, we exactly calculate the average density profile $\langle n_x\rangle_t$ 
for the domain wall and the alternating initial conditions by using the integral formula, Eq.~\eqref{eq:integral}. 
In both cases, we determine the asymptotic form of $\langle n_x\rangle_t$ for large $t$.
For the domain wall initial condition, we show that the asymptotic form follows the diffusive scaling.
For the alternating initial condition, we analytically demonstrate the dynamical transition in the relaxation of $\langle n_x\rangle_t$
which was numerically found in Ref.~\cite{Haga_quasi}.
%----------------------------------------------------------

\subsection{Domain wall initial condition}
We will obtain the following exact expression of the average density for the domain wall initial condition,
\begin{equation}
\langle n_x\rangle_t = 2\int_0^t du \;e^{-4\gamma t(1-\sqrt{1-(u/t)^2})} J_{x-1}(2u) J_x(2u)~~~ \text{for}~x\geq 1,\label{eq:density_dw}
\end{equation}
where $J_x(u)$ is the Bessel function of the first kind. 
The exact solution for $x\leq 0$ is given from the relation $\braket{n_x}_t = 1- \braket{n_{1-x}}_t$, together
with Eq.~\eqref{eq:density_dw}.

To derive this, we first find a formula for the two-point correlation function for the domain wall initial condition. 
One first sees, by substituting Eq.~\eqref{eq:integral} into the first equation in Eq.~\eqref{eq:expansion_alt} with Eq.~\eqref{eq:green}, 
that it can be expressed as
\begin{equation}
G^{\mathrm{DW}}_{x_1,x_2}(t) =\sum_{y\leq 0} (-1)^{x_1-y} \oint_{|z_1|=r}\frac{dz_1}{2\pi i z_1}\oint_{|z_2|=1}\frac{dz_2}{2\pi i z_2}
						(z_1z_2)^{-y} e^{-iE(z_1,z_2)t}\phi(x_1,x_2;\downarrow,\uparrow|z_1,z_2).\label{eq:twopoint_dw_1}
\end{equation} 
Taking the geometric series in Eq.~\eqref{eq:twopoint_dw_1}, which converges since $|z_1|=r\ll 1$, and 
substituting Eq.~\eqref{eq:vector_bethe} into Eq.~\eqref{eq:twopoint_dw_1}, we get 
\begin{equation}
G^{\mathrm{DW}}_{x_1,x_2}(t)=
(-1)^{x_1}\oint_{|z_1|=r}\frac{dz_1}{2\pi i z_1}\oint_{|z_2|=1} \frac{dz_2}{2\pi i z_2} e^{-iE(z_1,z_2)t} z_2^{x_{Q(1)}}
z_1^{x_{Q(2)}} \frac{s_1-s_2}{s_1-s_2-2\gamma}\frac{1}{1+z_1z_2} 
.\label{eq:twopoint_domain}
\end{equation}
Due to the particle-hole symmetry and the inversion symmetry, it holds that
\begin{equation}
G^{\mathrm{DW}}_{x_1,x_2}(t) = \delta_{x_1,x_2}-(-1)^{x_1+x_2} G^{\mathrm{DW}}_{1-x_2,1-x_1}(t).
\end{equation}
Hence, in the following, we assume that $x_{Q(2)}\geq 1$ without loss of generality.
After some algebra, as detailed in Appendix~\ref{app:derivation_dw}, the two-point correlation function can be expressed as
\begin{equation}
\begin{split}
G^{\mathrm{DW}}_{x_1,x_2}(t) &=  (-i)^{x_1}i^{x_2}  \int_0^{t} \frac{du}{\alpha(u)} e^{-4\gamma t (1- \alpha(u))} 
\left(\frac{u/t}{\alpha(u)+1}\right)^{|x_2-x_1|}\\
&\times \left[(\alpha(u)+1)J_{x_{Q(1)}}(2u)J_{x_{Q(2)}-1}(2u)+ (\alpha(u)-1)J_{x_{Q(1)}-1}(2u)J_{x_{Q(2)}}(2u)\right]
\end{split}\label{eq:exact_twopoint_dw}
\end{equation}
with $\alpha(u):= \sqrt{1-(u/t)^2}$.
Finally by setting $x_1=x_2(=x)$, we arrive at the exact solution of $\langle n_x \rangle_t$, Eq.~\eqref{eq:density_dw}.

Note that the $\gamma=0$ case of Eq.~\eqref{eq:density_dw} corresponds to the exact solution for the XX spin chain without dissipation\cite{Antal_xx},
since one can map the one-dimensional free fermions to the XX spin chain by the Jordan-Wigner transformation
\cite{Coleman_introduction}. 
The detail for this is described in Appendix~\ref{app:limit_dw}.
On the other hand, as $\gamma \to \infty$, it approaches the exact solution for SEP\cite{Liggett_interacting,Spitzer_interaction,Schutz_non},
as we discuss in the next section. 

From Eq.~\eqref{eq:density_dw}, one can further derive an exact formula of an average integrated current $\langle Q\rangle_t$,
which is defined as the mean of the total current from time 0 to $t$ across the bond between the site 0 and the site 1, but is
simply given by $\langle Q\rangle_t = \sum_{x\geq1} \langle n_x \rangle_t$ for the domain wall initial condition. 
Using Eq.~\eqref{eq:density_dw} and the following formula for the Bessel function,
\begin{equation}
\sum_{k=0}^\infty J_{m+k}(t)J_{n+k}(t)=\frac{t}{2(m-n)}[J_{m-1}(t)J_n(t)-J_{m}(t)J_{n-1}(t)],\label{eq:bessel_kernel}
\end{equation}
we have the exact solution of the average integrated current,
\begin{equation}
\langle Q \rangle_t  =2 \int_0^{t} du \;e^{-4\gamma t(1-\sqrt{1-(u/t)^2})} u( J_0(2u)^2 +J_1(2u)^2).\label{eq:current_dw}
\end{equation}
The derivation of Eq.~\eqref{eq:bessel_kernel} is similar to that of the Christoffel-Darboux formula
for orthogonal polynomials\cite{Szeg_orthogonal}, and is omitted here.

%-------------------------------------------------------------
We next determine the asymptotic behavior of $\langle n_x \rangle_t$ and $\langle Q\rangle_t$ for large $t$. 
To state the results, it is useful to
introduce the rescaled time $\tau := t/2\gamma$ and the rescaled coordinate $X:=x/\sqrt{\tau}$.
For the density $\langle n_x \rangle_t$, its asymptotic form up to $\mathcal{O}(1/\tau)$ is given by
\begin{equation}
\langle n_x \rangle_t \simeq \frac{1}{\sqrt{\pi}} \int^{\infty}_{X/2} du\;  e^{-u^2} + \frac{e^{-X^2/4} }{4\sqrt{\pi \tau}}
-\frac{e^{-X^2/4}}{96\sqrt{\pi}\tau}\left[(1+\frac{3}{4\gamma^2}) X -(1-\frac{3}{4\gamma^2})\frac{X^3}{2}\right].
\label{eq:asymptotics_dw}
\end{equation}
The asymptotic form of $\langle Q\rangle_t$ up to $\mathcal{O}(1/\sqrt{\tau})$ reads
\begin{equation}
\langle Q\rangle_t \simeq \sqrt{\frac{\tau}{\pi}} -\frac{1}{16\sqrt{\pi \tau}} \left[1+\frac{3}{4\gamma^2}\right].
\label{eq:asymptotics_current}
\end{equation}
The derivation of them are given in Appendix \ref{app:asymptotics_dw}, where asymptotic behaviors of the off-diagonal element in Eq.~\eqref{eq:twopoint_domain} 
is also derived. 

\begin{figure}[htbp]
    \centering
    \includegraphics[width=0.6\textwidth]{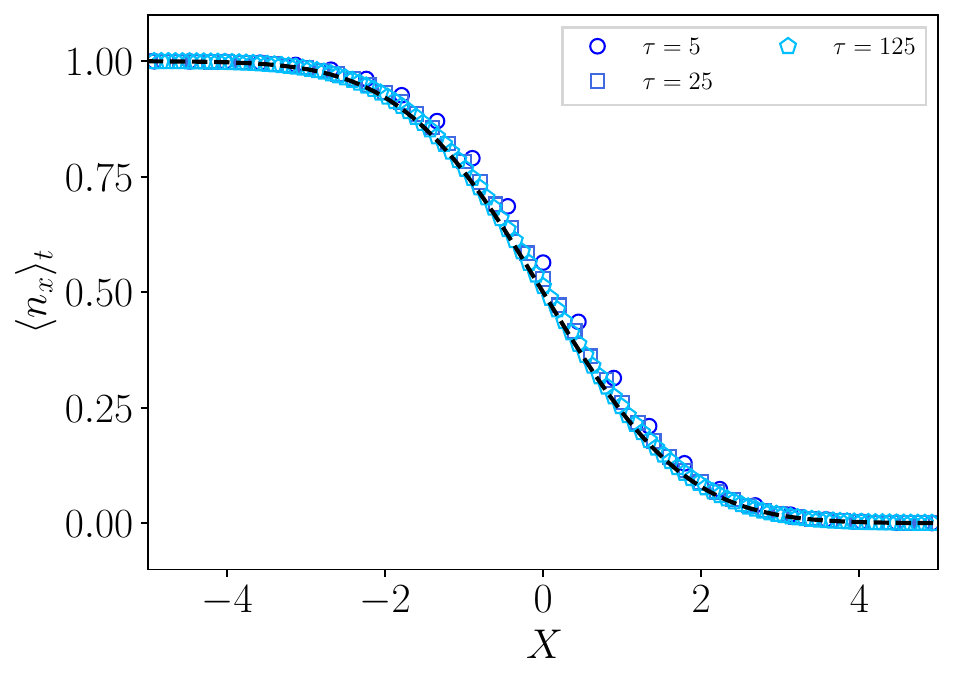} 
    \caption{Density profile $\langle n_x\rangle_t$ with rescaled coordinate $X =x/\sqrt{\tau}$. 
    The colored markers show the exact solution \eqref{eq:density_dw} for $\gamma =1$,
    while the dased line does the leading term of the asymptotic expansion \eqref{eq:asymptotics_dw}.}
    \label{fig:density_dw} 
\end{figure}

The leading terms of Eqs.~\eqref{eq:asymptotics_dw} and \eqref{eq:asymptotics_current} illustrate 
the late time dynamics becomes diffusive.
In Fig.~\ref{fig:density_dw}, we show that the exact solution of $\langle n_x\rangle_t$, Eq.~\eqref{eq:density_dw}
and the leading term of Eq.~\eqref{eq:asymptotics_dw}.
One clearly finds that $\langle n_x\rangle_t$ obeys the diffusive scaling from Fig.~\ref{fig:density_dw}.
The fact that the system becomes diffusive
has been previously discussed using a spectral analysis\cite{Esposito_emergence,Esposito_exactly}, 
a hydrodynamic argument\cite{Carollo_fluctuating}, a renormalization-group analysis\cite{Fujimoto_impact},
and an analysis of non-equilibrium steady states\cite{Znidaric_exact,Znidaric_large,Turkeshi_diffusion}.
Our derivation, relying on our exact solution, gives a firm confirmation of this fundamental diffusive transport.
Furthermore, our formula will allow us to compare the asymptotic forms with those for SEP and evaluate corrections from SEP in the next section.
%-------------------------------------------------------------------------------------
%------------------------------------------------------------------------------------
 
%-------------------------------------------------------------
\subsection{Alternating initial condition}
As in the case of the domain wall initial condition, 
we have the exact solution of $G^{\mathrm{Alt}}_{x_1,x_2}(t)$ as follows,
\begin{equation}
\begin{split}
G^{\mathrm{Alt}}_{x_1,x_2}(t)&=\frac{1+(-1)^{x_1}}{2}\delta_{x_1,x_2}
 + (-1)^{x_{Q(1)}} i^{x_2-x_1}
\int_0^{t}\frac{du}{\alpha(u)}e^{-4\gamma t(1-\alpha(u))}
\\ &\times
[(\alpha(u)+1)J_{|x_2-x_1|-1}(4u) -(\alpha(u)-1)J_{|x_2-x_1|+1}(4u)]\left(\frac{u/t}{1+\alpha(u)}\right)^{|x_2-x_1|}\label{eq:exact_twopoint_alt}
\end{split}
\end{equation}
with $\alpha(u)= \sqrt{1-(u/t)^2}$.
The derivation of Eq.~\eqref{eq:exact_twopoint_alt} is given in Appendix \ref{app:derivation_alt}. 
In Eq.~\eqref{eq:exact_twopoint_alt}, the case $x_1=x_2(=x)$ provides the exact solution of $\langle n_x \rangle_t$,
\begin{equation}
\langle n_x\rangle_t = \frac{1+(-1)^{x}}{2} + \frac{(-1)^{x-1}}{2} \int_0^{4t} du \;e^{-4\gamma t(1-\sqrt{1-(u/4t)^2})} J_1(u).\label{eq:density_alt}
\end{equation}

\begin{figure}[htbp]
    \centering
    \includegraphics[width=0.6\textwidth]{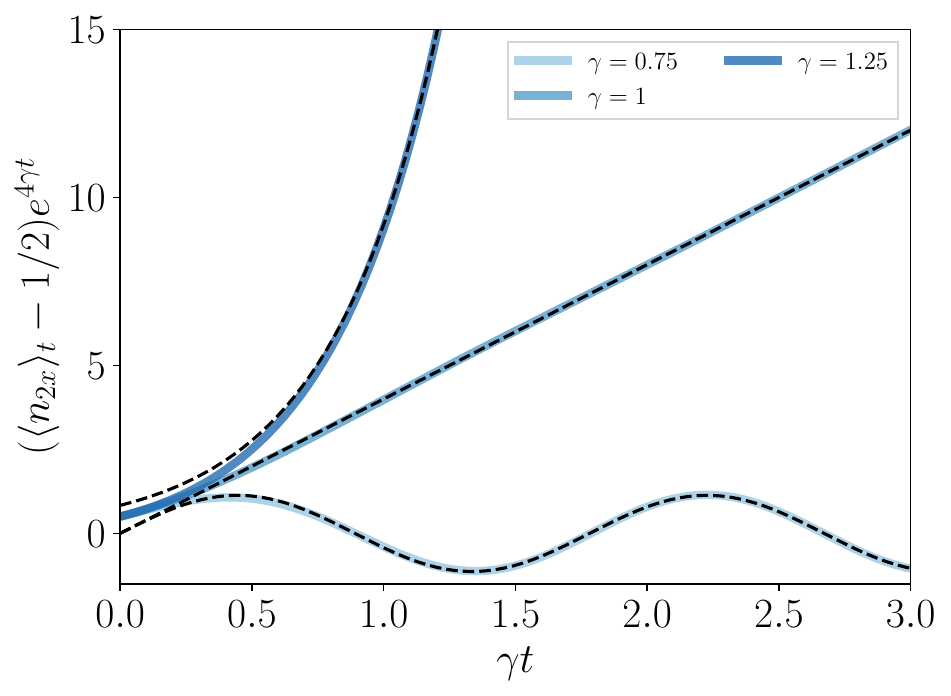} 
    \caption{Time evolution of $\langle n_x \rangle$ for the alternating initial condition.
    The colored lines show the exact solution, Eq.~\eqref{eq:density_alt}.
    The dashed lines show the asymptotic form, Eq.~\eqref{eq:density_asymptotics_alt}.}
    \label{fig:density_alt} 
\end{figure}

Using the exact solution, we obtain the asymptotic form of $G^{\mathrm{Alt}}_{x_1,x_2}(t)$
for large $t$.
Specifically, the density $\langle n_x \rangle_t$ for $t\gg1$ becomes
\begin{equation}
\langle n_x \rangle_t \simeq 
\begin{cases}
\frac{1}{2} + \frac{(-1)^x\gamma}{2\sqrt{\gamma^2-1}}e^{-4\gamma t(1-\sqrt{1-1/\gamma^2})} 
&\text{for} \;\; \gamma >1,
\\
\frac{1}{2} + 4(-1)^x t e^{-4 t} &\text{for}\;\; \gamma =1,
\\
\frac{1}{2}+ \frac{(-1)^x \gamma}{\sqrt{1-\gamma^2}}\sin[4t\sqrt{1-\gamma^2}] e^{-4\gamma t}
&\text{for} \;\; 0< \gamma <1.
\end{cases}\label{eq:density_asymptotics_alt}
\end{equation}
See Appendix \ref{app:asymptotics_alt} for the derivation and the asymptotic form of the off-diagonal element.
The equation~\eqref{eq:density_asymptotics_alt} illustrates
that the relaxation of $\langle n_x\rangle_t$ undergoes the dynamical transition from oscillatory decay to
over-damped decay as $\gamma$ increases (see Fig.~\ref{fig:density_alt}).
%-------------------------------------------------------------------------------------
%-------------------------------------------------------------------------------------
This dynamical transition was previously identified through the numerical analysis
in Ref.~\cite{Haga_quasi}.
The equation~\eqref{eq:density_asymptotics_alt} gives the analytical demonstration for the presence of the transition.
Moreover, in Appnedix \ref{app:asymptotics_alt}, we show that the off-diagonal element 
of $G^{\mathrm{Alt}}_{x_1,x_2}(t)$ also exhibits the similar dynamical transition.

Before closing this subsection, we comment on the polynomial correction \eqref{eq:density_asymptotics_alt}
to the exponential decay at the transition point $\gamma=1$.
Similar polynomial corrections are known to appear in the non-hermitian systems at exceptional points
\cite{Ashida_non,Minganti_quantum,Bergholtz_exceptional,Heiss_physics,Ganainy_non},
which are special parameter values where both eigenvalues and eigenvectors coalesce.
As demonstrated in the derivation in Appendix~\ref{app:asymptotics_alt}, 
the polynomial correction arises from the coalescence of the poles at $\gamma=1$ in the contour integral, 
and thus the phenomenon is analogous to the coalescence of eigenvectors at exceptional points.
However, the relation between exceptional points and the coalescence of the poles
is currently unclear
because the Liouvillian of our model cannot be reduced to the finite-dimensional eigenvalue problem.
We leave this issue for future work.

%------------------------------------------------------------------------------------------------------------------------
\section{Comparison to SEP}\label{sec:correction}
Several studies have discussed connections between our model and a well-known classical Markov process, SEP\cite{Cai_algebraic, Znidaric_large, Carollo_fluctuating}, 
in which many particles perform symmetric random jumps to nearest neighbor sites on a one-dimensional lattice with volume exclusion
\cite{Liggett_interacting,Spitzer_interaction,Schutz_non}. 
However the arguments on whether the time evolution of our systems in the long time limit is equivalent to that of SEP are not rigorous.
In this section, we confirm this for the density profile by using the exact results obtained in the previous section and identify corrections from SEP.

\subsection{Background and motivation of this  section}
As mentioned in the previous section,
it is known that, in the strong dephasing limit, our system can be effectively described by SEP.
According to Ref.~\cite{Cai_algebraic}, at the level of the second-order perturbation theory, 
the equation of motion for the diagonal element of the density matrix 
$P(x_1,\cdots,x_N;t):= \langle 0\vert a_{x_1}\cdots a_{x_N} \rho(t) a^{\dagger}_{x_N}\cdots a^\dagger_{x_1}\ket{0} $ can be written 
in terms of the spin-1/2 operators as
\eq{
\frac{d}{dt} \vert P(t)\rangle  \simeq -\frac{1}{2\gamma} \sum_{x\in \mathbb{Z}}
\left[\frac{1}{2}(s^+_x s^-_{x+1} + s^+_{x+1}s^-_x )+s_x^z s_{x+1}^z-\frac{1}{4} \right]
 \vert P(t)\rangle
 \label{eq:peturbation}
}
where we define $\vert P(t)\rangle:= \sum_{x_1< \cdots< x_N}P(x_1,\cdots,x_N;t) s^+_{x_1}\cdots s^+_{x_N} \ket{0}_{\mrm{spin}}$
denoting all the down-spin state as $\ket{0}_{\mrm{spin}}$.
Here, $s_x^+$, $s_x^-$, and $s^z_x$ represent the raising, the lowering, and the $z$-spin operators at site $x$, respectively.
After the time rescaling $\tau=t/2\gamma$, the above equation is identical to the master equation for SEP\cite{Liggett_interacting,Spitzer_interaction,Schutz_non}.

There have also been some arguments, suggesting that the late-time dynamics becomes equivalent to that in SEP
even for a fixed and finite $\gamma$. For example in Ref.~\cite{Carollo_fluctuating} the authors approximate the dynamics 
by a classical fluctuating hydrodynamics and observed the appearance of the one for SEP. In Ref.~\cite{Znidaric_large} 
the diffusion constant $D$ and mobility $\sigma$ were calculated to be the same as the ones 
for SEP by studying the non-equilibrium steady states.  
However, whether the long time behaviors of our model are actually not modified by the higher-order perturbation terms
and do agree with those of SEP is a nontrivial question.  

In this section, we verify these connections to SEP for the density profile under both the domain wall and the alternating initial conditions
by using the exact results obtained in the previous section.
We first consider the strong dephasing limit of the exact solutions for the density and check the consistency with the second-order perturbation theory.
Next order terms are also obtained for both initial conditions.
In the long time limit, while the leading terms of asymptotic behaviors agree with those of SEP, differences emerge 
between the two initial conditions.
For the domain wall initial condition, the average density matches that of SEP up to the subleading term,
whereas for the alternating initial conditions, it only does up to the leading term.
Furthermore, we find that the transition behavior observed in our system for the alternating initial condition 
(see Eq.~\eqref{eq:density_asymptotics_alt}) is absent in SEP.

Before going to the detailed analysis, we give the time evolution of the density $\braket{n_x}^{\mrm{SEP}}_\tau$ in SEP.
The time evolution of $\langle n_x \rangle^{\mathrm{SEP}}_\tau$ follows simply the discretized diffusion equation
\cite{Liggett_interacting,Spitzer_interaction,Schutz_non},
\begin{equation}
\frac{d}{d\tau}\langle n_x \rangle^{\mathrm{SEP}}_\tau = 
\langle n_{x+1} \rangle^{\mathrm{SEP}}_{\tau} + \langle n_{x-1} \rangle^{\mathrm{SEP}}_{\tau} -2 \langle n_x \rangle^{\mathrm{SEP}}_\tau.
\label{eq:timeevo_sep}
\end{equation}
We compare $\braket{n_x}_t$ at the rescaled time $\tau=t/2\gamma$ with $\braket{n_x}^{\mrm{SEP}}_\tau$ in the following section.
Note that this time rescaling is appropriate for the comparison when $\gamma \gg 1$ as indicated by Eq.~\eqref{eq:peturbation}.
Even for finite $\gamma$, the previous studies in Refs.~\cite{Znidaric_large, Carollo_fluctuating} 
suggest this rescaling is appropriate, and indeed it is already used in Eqs.~\eqref{eq:asymptotics_dw} and \eqref{eq:asymptotics_current} 
in the previous section.

%----------------------------------------------------------------------------------------------------------------
\subsection{Domain wall initial condition}
The density of SEP with the domain wall initial condition is easily found by solving Eq.~\eqref{eq:timeevo_sep} to be
\begin{equation}
\langle n_x \rangle^{\mathrm{SEP}}_\tau = \oint_{|\eta|<1}\frac{d\eta}{2\pi i\eta} e^{\tau(\eta+1/\eta-2)}\frac{\eta^x}{1-\eta}.
\end{equation} 
Its asymptotic form for long time, i.e., for $\tau\gg1$, is given by 
\begin{equation}
\langle n_x \rangle^{\mathrm{SEP}}_\tau  \simeq \frac{1}{\sqrt{\pi}}\int^{\infty}_{X/2} e^{-u^2}du + \frac{e^{-X^2/4}}{4\sqrt{\pi\tau}} 
+\frac{e^{-X^2/4}}{96\sqrt{\pi}\tau} [X^3/2-X]\label{eq:asymptotics_sep}
\end{equation}
with the rescaled coordinate $X=x/\sqrt{\tau}$.
The derivation is similar to Eq.~\eqref{eq:asymptotics_dw}, using a saddle point method around $\eta=1$. 
In the following, we compare the density $\braket{n_x}_t$ in our model for large $\gamma$ (with finite $\tau$) and for large $\tau$ (with finite $\gamma$)
with the density $\braket{n_x}^{\mrm{SEP}}_\tau$ in SEP under the domain wall initial condition.

First the large $\gamma$ (with finite $\tau$) behavior of the density $\braket{n_x}_t$   in Eq. \eqref{eq:density_dw}  is found to be
\eq{
\braket{n_x}_t \simeq & \oint_{|\eta|<1} \frac{d\eta}{2\pi i \eta} \frac{\eta^x}{1-\eta} e^{\tau(\eta+1/\eta-2)}\\
-&\frac{1}{16\gamma^2} \oint \frac{d\eta}{2\pi i \eta} \eta^{x-2}e^{\tau(\eta+1/\eta-2)} [\tau(1-\eta)^3+2\eta(1-\eta)],
\label{eq:zeno_dw}
}
where the approximation error is bounded by $\mathcal{O}(\gamma^{-4})$. This is derived in Appendix \ref{app:zeno}. 
Note that the first term is equivalent to $\braket{n_x}^{\mrm{SEP}}_\tau$.
This is consistent with the known fact that, in the strong dephasing limit, the Liouvillian of the tight binding chain with the dephasing is 
approximated as the generator of SEP as a result of the second-order perturbation\cite{Cai_algebraic}. 
From our exact formula, it is rather straightforward to find the next order contribution, nemely the second term in Eq.~\eqref{eq:zeno_dw}. 
It would be possible and interesting to derive the same term by considering higher orders of perturbation of the Liouvillian, although it might require 
a fairly sophisticated method like a singular perturbation theory
\cite{Chen_renormalization_1,Chen_renormalization_2,
Kunihiro_geometrical,Kunihiro_renormalization,Fujimoto_impact,Longhi_dephasing,Dhawan_anomalous}. 

Next, we consider the large $\tau$ (with finite $\gamma$) behavior. Such asymptotic expressions have already been derived in 
Eq.~\eqref{eq:asymptotics_dw} for our model and in Eq.~\eqref{eq:asymptotics_sep} for SEP. 
Comparing these two, one observes that the asymptotic form of $\langle n_x\rangle_t$ matches that of
$\langle n_x\rangle^{\mathrm{SEP}}_\tau$, up to the subleading term $\mathcal{O}(1/\sqrt{\tau})$. 
Moreover the correction from SEP, defined as $\Delta \langle n_x\rangle_t := \langle n_x \rangle_t -\langle n_x\rangle^{\mathrm{SEP}}_\tau$, 
can be calculated and is given by
\begin{equation}
\Delta \langle n_x\rangle_t  \simeq -\frac{e^{-X^2/4}}{128\gamma^2 \sqrt{\pi}\tau}(X^3/2+X).
\label{eq:density_correction}
\end{equation}
Note that the correction $\langle n_x\rangle_t$ is of order $1/\tau$ relative to the leading term.
One may have noticed that the large $\tau$ (with finite $\gamma$) expansion in Eq.~\eqref{eq:contour_dw} in Appendix~\ref{app:asymptotics_dw} (which gives Eq.~\eqref{eq:asymptotics_dw}) is in fact exactly the same as 
the large $\gamma$ (with finite $\tau$) expansion in Eq.~\eqref{eq:zeno_dw}.
The fact that the long time asymptotic expression agrees with the strong dissipation expansion is a nontrivial observation based on our exact formulas, although a posteriori it could be understand as a consequence of properties of the spectrum and the eigenstate of the 
two-particle Fermi-Hubbard Hamiltonian.

Similarly, one can evaluate the correction for the average integrated current $\langle Q\rangle_t$.
In SEP, the average integrated current $\langle Q\rangle^{\mathrm{SEP}}_\tau$
can be expressed by the modified Bessel function $I_n(\tau)$ as
\begin{equation}
\langle Q\rangle^{\mathrm{SEP}}_\tau = \sum_{x\geq 1}\langle n_x\rangle^{\mathrm{SEP}}_\tau= \tau e^{-2\tau}[I_0(2\tau)+I_1(\tau)].
\end{equation}
The asymptotic expansion of the modified Bessel function\cite{Math_formula} yields the asymptotic form
of $\langle Q\rangle^{\mathrm{SEP}}_\tau$,
\begin{equation}
\langle Q\rangle^{\mathrm{SEP}}_\tau \simeq \sqrt{\frac{\tau}{\pi}} -\frac{1}{16\sqrt{\pi\tau} }.
\label{eq:asymptotics_current_sep}
\end{equation}
Comparing Eq.~\eqref{eq:asymptotics_current} with Eq.~\eqref{eq:asymptotics_current_sep},
we have the correction $ \Delta \langle Q\rangle_t :=\langle Q\rangle_t -\langle Q\rangle^{\mathrm{SEP}}_\tau$,
\begin{equation}
\Delta \langle Q\rangle_t  \simeq -\frac{3}{64\gamma^2\sqrt{\pi \tau}}.
\end{equation} 
Similarly to $\Delta \langle n_x\rangle_t$, $\Delta \langle Q\rangle_t$ is of order $1/\tau$ relative to the leading term.
 
\begin{figure}[htbp]
    \centering
    \includegraphics[width=0.6\textwidth]{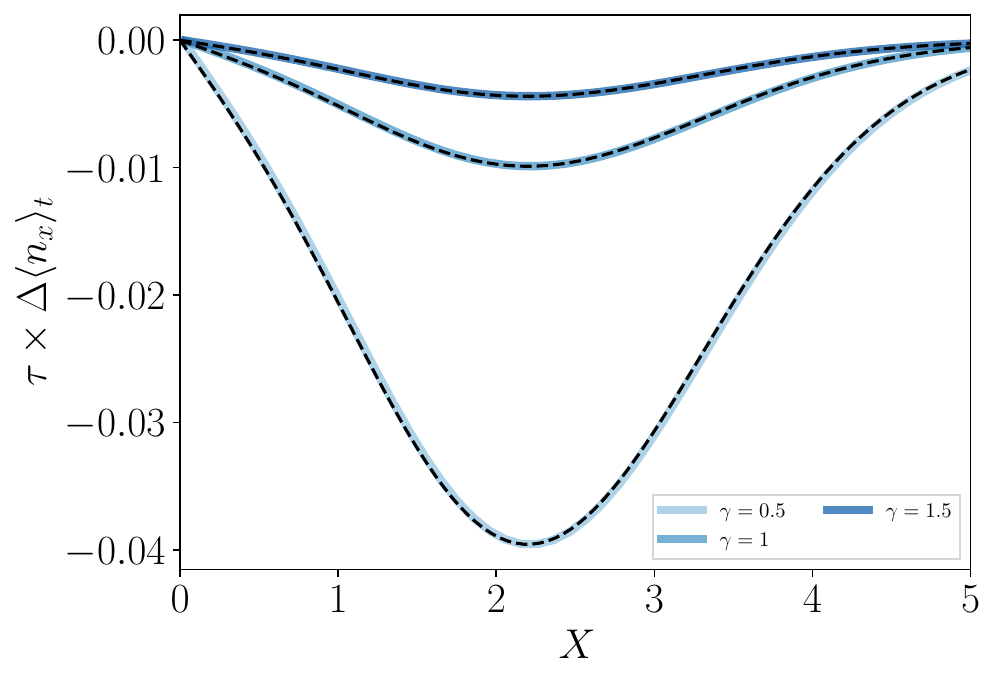} 
    \caption{Correction $\Delta \langle n_x\rangle_t$ for $\gamma=0.5,1$, and $1.5$. 
    The colored lines show the exact expression of $\Delta \langle n_x\rangle_t$ at $\tau =1000$.
    The dashed lines show the asymptotic form, Eq.~\eqref{eq:density_correction}.}
    \label{fig:density_correction} 
\end{figure}

In Fig.~\ref{fig:density_correction}, we show $\Delta \langle n_x\rangle_t$
and its asymptotic form, Eq.~\eqref{eq:density_correction} as functions of $X$.
Note that the asymptotic form, Eq.~\eqref{eq:density_correction} closely matches the exact expression.
As illustrated in Fig.~\ref{fig:density_correction},  
$\Delta \langle n_x\rangle_t $ becomes larger as $\gamma$ decreases, and this deviation is attributed to
the effect of the coherent hopping, which becomes more significant compared to the dissipation.
%-------------------------------------------------------------------------------------
%-------------------------------------------------------------------------------------

%-------------------------------------------------------------------------------------
\subsection{Alternating initial condition}
The density of SEP with the alternating initial condition is again easily found by solving Eq.~\eqref{eq:timeevo_sep} to be
\begin{equation}
\langle n_x \rangle^{\mathrm{SEP}}_\tau= \frac{1}{2} + \frac{(-1)^x}{2} e^{-4\tau}.
\label{eq:sep_alt}
\end{equation}
Similarly to the previous subsection, we compare $\braket{n_x}_t$ for
large $\gamma$ (with finite $\tau$) and large $\tau$ (with finite $\gamma$) with $\braket{n_x}^{\mrm{SEP}}_\tau$
under the alternating initial condition.

As derived in Appendix~\ref{app:zeno}, 
the large $\gamma$ (with finite $\tau$) behavior of $\braket{n_x}_t$ in Eq.~\eqref{eq:density_alt} is found to be
\eq{
\braket{n_x}_t = \frac{1}{2} + \frac{(-1)^x}{2} e^{-4\tau} \left[ 1 +\frac{1-2\tau}{2\gamma^2}\right] + \order{\gamma^{-4} }.
\label{eq:zeno_alt}
}
From the above equation, one can confirm 
that the leading term is given by $\braket{n_x}^{\mrm{SEP}}_\tau$ for the alternating initial condition as well.

We next consider the large $\tau$ (with finite $\tau$) case.
From Eq.~\eqref{eq:density_asymptotics_alt}, the large $\tau$ behavior of $\braket{n_x}_t$ can be written as
\begin{equation}
\langle n_x \rangle_t \simeq 
\begin{cases}
\frac{1}{2} + \frac{(-1)^x\gamma}{2\sqrt{\gamma^2-1}}e^{-8\gamma^2 \tau(1-\sqrt{1-1/\gamma^2})} 
&\text{for} \;\; \gamma >1,
\\
\frac{1}{2} + 8(-1)^x \tau e^{-8\tau} &\text{for}\;\; \gamma =1,
\\
\frac{1}{2}+ \frac{(-1)^x \gamma}{\sqrt{1-\gamma^2}}\sin[8\gamma\tau\sqrt{1-\gamma^2}] e^{-8\gamma^2 \tau}
&\text{for} \;\; 0< \gamma <1.
\end{cases}\label{eq:density_asymptotics_alt_rescaled}
\end{equation}
Comparing Eq.~\eqref{eq:density_asymptotics_alt_rescaled} with Eq.~\eqref{eq:sep_alt},
one finds that the leading term takes the same value of $1/2$ in both cases, but the subleading term, namely 
the exponential decay to the stationary value, differs.
Specifically, the transition behavior is absent in $\langle n_x\rangle^{\mathrm{SEP}}_\tau$, indicating the relaxation behavior is qualitatively different
between the two models.

This difference is due to the existence of a translational symmetry by two lattice units, i.e., 
$\braket{n_{x+2}}_t=\braket{n_x}_t$ and $\braket{n_{x+2}}^{\mrm{SEP}}_\tau = \braket{n_x}^{\mrm{SEP}}_\tau$, for the alternating initial condition. 
This symmetry restricts the Fourier components of the average density to the wavenumbers $k=0$ and $k=\pi$.
Since the Fourier component at $k=0$ corresponds to the stationary value and is time-independent in both cases, 
the dynamics of the average density is determined solely by the component at $k=\pi$.
From a physical point of view, the correspondence with SEP is expected to hold on long distance $1/k \gg 1$ and long time scales $\tau \gg 1$.
Therefore, it is not surprising that the time dependence of the Fourier component at $k=\pi$ differs between the two models.
In contrast, the domain wall initial condition does not have the translational symmetry, resulting in the identical relaxation behavior between the two models.

%-------------------------------------------------------------------------------------------------------------
\section{Conclusion and future prospect}
In this work, we studied the many-body dynamics of the tight-binding chain with the dephasing noise on the infinite interval. 
Our first result is the exact formula of the Green's function for the two-point correlation function expressed as the double 
contour integral, Eq.~\eqref{eq:integral}. It has been derived by relating it to the two particle problem of the 
Fermi-Hubbard model with imaginary interaction and subsequently employing the Bethe ansatz techniques.
In the derivation, we could avoid solving the Bethe equations because we considered the infinite system from the outset. 
A similar approach has been successfully applied in the context of classical stochastic interacting systems.
To the best of our knowledge, this is the first example of such applications to open quantum systems. 

Our second results are the exact solution of the average density for the domain wall initial condition, Eq.~\eqref{eq:density_dw} and for the alternating initial condition, Eq.~\eqref{eq:density_alt}.
They were obtained by summing the Green's function over these two initial conditions. 
We also studied their long time behaviors.  
For the domain wall initial condition, we determined the large $t$ asymptotic form up to order $1/t$, 
and showed that it obeys the diffusive scaling.
For the alternating initial condition, we demonstrated
the relaxation of the average density exhibits the dynamical transition from oscillatory decay to
over-damped decay as $\gamma$ increases.

Our exact results are valid for any finite values of the dephasing strength $\gamma$ and time $t$, giving precise confirmations of 
the previous findings for the limiting behaviors.  
It has been known that the strong dephasing limit of our system is described by a classical 
Markov chain known as the symmetric simple exclusion process\cite{Cai_algebraic}. 
Our formulas confirmed this for the density profiles starting from the two initial conditions. 
They also lead to the systematic expansions of the density for large $\gamma$.
It may be compared with the higher-order perturbation theory, which can be applied even for 
non Bethe ansatz solvable models.
There are also a few literature\cite{Znidaric_large,Carollo_fluctuating}, suggesting that, even for a finite strength of the dephasing, the long time behaviors of the model 
would agree with those of SEP. 
The asymptotic behaviors of our exact formulas allow us to verify this. 
The leading terms of our formulas do agree with those of SEP. 
The corrections to the behaviors of SEP can also be studied from our formulas. 
We observed that a symmetry of the initial condition may change the long time behaviors and also that 
the long time behaviors may be understood as results of large depahsing behaviors. 
These analyses would be useful for understanding effects of the dephasing in open quantum systems in general. 

In this work, we focused on the calculation of the density profile as the simplest physical quantity. 
One of the most natural and intriguing directions for extending our methods is to study the system's fluctuations. 
For example, for the fluctuation of the integrated current, exact results for non-equilibrium quantum many-body systems
are available only for systems which can be reduced to free fermions\cite{Antal_logarithmic,Moriya_exact}, and 
going beyond free systems would be a critically important challenge. 
We plan to address this problem in our next work. 

There are several types of Bethe ansatz solvable models for the GKSL equation
\cite{Medvedyeva_exact, Rubio_integrability, Claeys_dissipative, Ziolkowska_yang, De_constructing, Essler_integrability, 
Nakagawa_exact, Buca_XXZ, Rrobertson_exact, Ekman_liouvillian,Shibata_ashkin,Yamamoto_universal,Yamamoto_universal_SU(N)}.  
It would be fascinating to investigate many-body dynamics of these models by generalizing the methods in this paper of employing the Bethe ansatz directly in the infinite lattice, and we believe that our approach would provide a useful tool to uncover various interesting non-equilibrium properties. 

\vskip 0.666cm
\noindent 
\textbf{Acknowledgments.}
The authors are grateful to Ryo Hanai, Masaya Nakagawa, Yuma Nakanishi, Kazuho Suzuki, Kazuki Yamamoto, and Hironobu Yoshida
for helpful discussions and comments.
The work of KF has been supported by JSPS KAKENHI Grant No. JP23K13029. 
The work of TS has been supported by JSPS KAKENHI Grants No. JP21H04432, No. JP22H01143, and No. JP23K22414.

%-------------------------------------------------------------------------------------------------------------
\appendix

%-----------------------------------------------------------------------------------------------------
\section{Derivation of Eq.~(\ref{eq:twopoint_domain})}\label{app:derivation_dw}
We derive the exact solution of $G^{\mathrm{DW}}_{x_1,x_2}(t)$ with the domain wall initial condition.
Specifically, we show that Eq.~\eqref{eq:twopoint_domain}, namely
\begin{equation}
G^{\mathrm{DW}}_{x_1,x_2}(t)=
(-1)^{x_1}\oint_{|z_1|=r}\frac{dz_1}{2\pi i z_1}\oint_{|z_2|=1} \frac{dz_2}{2\pi i z_2} e^{-iE(z_1,z_2)t} z_2^{x_{Q(1)}}
z_1^{x_{Q(2)}} \frac{s_1-s_2}{s_1-s_2-2\gamma}\frac{1}{1+z_1z_2} 
\end{equation}
can be transformed to Eq.~\eqref{eq:exact_twopoint_dw}.
In the derivation, we assume that $x_{Q(2)}\geq 1$ without loss of generality as explained in the main text.

After the substitutions, $z_1\to iz_1$ and $z_2\to iz_2$, the above equation becomes
\begin{equation}
G^{\mathrm{DW}}_{x_1,x_2}(t)
= (-i)^{x_1}i^{x_2} \oint_{|z_1|=r}\frac{dz_1}{2\pi i z_1}\oint_{|z_2|=1} \frac{dz_2}{2\pi i z_2}
\frac{e^{t(-z_1+z_1^{-1}-z_2+z_2^{-1}-4\gamma)} }{2(c_1-c_2-2\gamma)}
z_2^{x_{Q(1)}-1}z_1^{x_{Q(2)}-1} (z_2-z_1).
\end{equation}
By inserting the following identity
\begin{equation}
\frac{1}{2(c_1-c_2-2\gamma)}= \int_0^{t} ds\;e^{2(c_2-c_1+2\gamma)s} -\frac{e^{2(c_2-c_1+2\gamma)t}}{2(c_2-c_1+2\gamma)},\label{eq:trick}
\end{equation}
we have
\begin{equation}
\begin{split}
G^{\mathrm{DW}}_{x_1,x_2}(t)
&=(-i)^{x_1} i^{x_2} \int_0^t ds \;e^{-4\gamma(t-s)} 
\oint_{|z_1|=r}\frac{dz_1}{2\pi i z_1}\oint_{|z_2|=1} \frac{dz_2}{2\pi i z_2}
\\ &\times
e^{-(t+s)z_1+(t-s)z_1^{-1}} e^{-(t-s)z_2+(t+s)z_2^{-1}}
z_2^{x_{Q(1)}-1}z_1^{x_{Q(2)}-1}  (z_2-z_1)
\\
&+ (-i)^{x_1} i^{x_2}\oint_{|z_1|=r}\frac{dz_1}{2\pi i z_1}\oint_{|z_2|=1} \frac{dz_2}{2\pi i z_2}
\frac{e^{2t(-z_1+z_2^{-1} )} }{2(c_1-c_2-2\gamma)}z_2^{x_{Q(1)}-1}z_1^{x_{Q(2)}-1} (z_2-z_1).
\end{split}
\end{equation}
The second term in the above equation is 0 since the integrand is holomorphic with respect to $z_1$
inside the $z_1$-contour.
Thus, we have
\begin{equation}
\begin{split}
G^{\mathrm{DW}}_{x_1,x_2}(t)
&=(-i)^{x_1} i^{x_2} \int_0^t ds \;e^{-4\gamma(t-s)} 
\oint \frac{dz_1}{2\pi i z_1}\oint \frac{dz_2}{2\pi i z_2}
\\ &\times
e^{-(t+s)z_1+(t-s)z_1^{-1}} e^{-(t-s)z_2+(t+s)z_2^{-1}}
z_2^{x_{Q(1)}-1}z_1^{x_{Q(2)}-1}  (z_2-z_1).
\end{split}\label{eq:derrida_trick}
\end{equation}
Note that the poles of $1/(c_1-c_2-2\gamma)$ are eliminated by inserting the identity, Eq.~\eqref{eq:trick}.
Similar calculations were used in Refs.~\cite{Derrida_current,Moriya_exact}.

Making the substitutions $z_1 \to \sqrt{(t-s)/(t+s)}z_1,\;z_2 \to \sqrt{(t+s)/(t-s)}z_2$ 
and using the contour integral representation of the Bessel function of the first kind $J_x(u)$\cite{Math_formula},
one finds that Eq.~\eqref{eq:derrida_trick} can be expressed in terms of $J_x(u)$ as
\begin{equation}
\begin{split}
G^{\mathrm{DW}}_{x_1,x_2}(t)
=(-i)^{x_1}i^{x_2} &\int_0^t ds \;e^{-4\gamma(t-s)} \left(\frac{t-s}{t+s}\right)^{|x_{2}-x_{1}|/2}\\
\times \biggl [  &\left(\frac{t-s}{t+s}\right)^{-1/2} J_{x_{Q(1)}}(2\sqrt{t^2-s^2})J_{x_{Q(2)}-1}(2\sqrt{t^2-s^2})\\
- &\left(\frac{t-s}{t+s}\right)^{1/2} J_{x_{Q(1)}-1}(2\sqrt{t^2-s^2})J_{x_{Q(2)}}(2\sqrt{t^2-s^2}) \biggr].\label{eq:exact_twopoint}
\end{split}
\end{equation}
Changing the variable as $s = \sqrt{t^2-u^2}$ yields Eq.~\eqref{eq:exact_twopoint_dw},
namely
\begin{equation}
\begin{split}
G^{\mathrm{DW}}_{x_1,x_2}(t) &=  (-1)^{x_1}i^{x_2}  \int_0^{t} \frac{du}{\alpha(u)} e^{-4\gamma t (1- \alpha(u))} 
\left(\frac{u/t}{\alpha(u)+1}\right)^{|x_2-x_1|}\\
&\times \left[(\alpha(u)+1)J_{x_{Q(1)}}(2u)J_{x_{Q(2)}-1}(2u)+ (\alpha(u)-1)J_{x_{Q(1)}-1}(2u)J_{x_{Q(2)}}(2u)\right]
\end{split}
\end{equation}
with $\alpha(u)= \sqrt{1-(u/t)^2}$.

%---------------------------------------------------------------------------------------------------------
\section{Zero dephasing limit of the average density}\label{app:limit_dw}
We consider the case $\gamma = 0$ in Eq.~\eqref{eq:density_dw} 
and check the consistency with the previous result\cite{Antal_xx}.
From Eq.~\eqref{eq:density_dw},
$\langle n_x \rangle_t$ for $\gamma = 0$ can be written as
\begin{align}
\langle n_x \rangle &= \int_0^{2t} du \; J_{x-1}(u)J_x(u)\\
                              &= \int_0^{2t}du\; 2J'_x(u) J_x(u) +J_x(u)J_{x+1}(u) \\
                              &= J^2_x(2t) + \langle n_{x+1} \rangle\\
                              &=\sum_{n \geq x} J^2_{n}(2t).
\end{align}
In the second line, we use the recursion relation $J_{x-1}(u)-J_{x+1}(u)=2J'_x(u)$\cite{Math_formula}.
This result is equivalent to Eq.~(16) of Ref. \cite{Antal_xx}.
%------------------------------------------------------------------------------------------------------------------------

%------------------------------------------------------------------------------------------------------------------------
\section{Asymptotic analysis for the domain wall initial condition}\label{app:asymptotics_dw}
In this section, we derive the asymptotic forms of $\langle n_x \rangle_t$, $\langle Q\rangle_t$, and $G^{\mathrm{DW}}_{x_1,x_2}(t)$ 
for large $\tau$.
We determine terms up to $\mathcal{O}(1/\tau)$ relative to a leading order 
for $\langle n_x\rangle_t$ and $\langle Q\rangle_t$, 
while we only determine a leading term for $G^{\mathrm{DW}}_{x_1,x_2}(t)$. 

We first derive the asymptotic form of $\langle n_x \rangle_t$ for large $\tau$, which is given by
\begin{equation}
\langle n_x \rangle_t \simeq \frac{1}{\sqrt{\pi}} \int^{\infty}_{X/2} du\;  e^{-u^2} + \frac{e^{-X^2/4} }{4\sqrt{\pi \tau}}
-\frac{e^{-X^2/4}}{96\sqrt{\pi}\tau}\left[(1+\frac{3}{4\gamma^2}) X -(1-\frac{3}{4\gamma^2})\frac{X^3}{2}\right].
\label{eq:asymptotics_dw_app}
\end{equation}
After the substitution $u\to \sqrt{\tau}u$,
Eq.~\eqref{eq:density_dw} becomes
\eq{
\langle n_x\rangle_t = 2\sqrt{\tau}\int_0^{2\gamma \sqrt{\tau}} du\; e^{-u^2} e^{-f(u)} J_{x-1}(2\sqrt{\tau}u)J_{x}(2\sqrt{\tau}u),
}
with 
\eq{
f(u):=u^2\frac{1-\sqrt{1-\frac{u^2}{4\gamma^2 \tau} }} {1+\sqrt{1-\frac{u^2}{4\gamma^2 \tau} }}.
}
As a first step in deriving the asymptotic form, we proceed to show the following inequality,
\eq{
\left| \braket{n_x}_t-2\sqrt{\tau}\int_0^{2\gamma\sqrt{\tau}}du\;
e^{-u^2}(1-\frac{u^4}{16\gamma^2\tau})J_{x-1}(2\sqrt{\tau}u)J_x(2\sqrt{\tau}u)\right|
\leq \frac{C}{\gamma^4 \tau^{3/2}},
\label{eq:inequality}
}
where $C$ is some numerical constant.
To show this, we first evaluate $|e^{-f(u)} -(1-\frac{u^4}{16\gamma^2\tau})|$ as follows,
\eqsnn{
\left| e^{-f(u)} -(1-\frac{u^4}{16\gamma^2\tau})\right| 
&\leq \left|e^{-f(u)}-(1-f(u))\right| + \left|\frac{u^4}{16\gamma^2}{\tau}-f(u)\right|
\\
&\leq  \frac{f^2(u)}{2} + \left|\frac{u^4}{16\gamma^2}{\tau}-f(u)\right|
\\
&=\frac{u^8}{32\gamma^4\tau^2}\frac{1}{(1+\sqrt{1-u^2/4\gamma^2\tau})^4} + \frac{u^4}{16\gamma^2\tau}\frac{u^2/4\gamma^2\tau +2(1-\sqrt{1-u^2/4\gamma^2\tau})}{(1+\sqrt{1-u^2/4\gamma^2\tau})^2}
\\
& \leq \frac{u^8}{32\gamma^4\tau^2} + \frac{3u^6}{64\gamma^4\tau^2} \leq \frac{u^6+u^8}{16\gamma^4\tau^2}.
}
Here, we use $|a+b| \leq |a| +|b|$ in the first line, $|e^{-x}-(1-x)|\leq x^2/2$ for $x\geq0$ in the second line, and $-x \leq -x^2$ for $0\leq x \leq 1$
in the fourth line.
From the above inequality and $|J_x(u)|\leq 1$ for $x\in \mathbb{Z}$,
the LHS of Eq.~\eqref{eq:inequality} can be evaluated as
\begin{align*}
\text{LHS of Eq.~\eqref{eq:inequality} } 
&\leq 2\sqrt{\tau}\int_0^{2\gamma \sqrt{\tau}}du\; e^{-u^2} \left| e^{-f(u)}-(1-\frac{u^4}{16\gamma^2\tau}) \right|
\\
&\leq \frac{1}{8\gamma^4\tau^{3/2}}\int_0^{2\gamma \sqrt{\tau}} du\; e^{-u^2} (u^6+u^8)
\\
&\leq \frac{1}{8\gamma^4\tau^{3/2}}\int_0^{\infty} du\; e^{-u^2} (u^6+u^8) 
= \frac{C}{\gamma^4\tau^{3/2}}.
\end{align*}
Thus, we obtain the inequality~\eqref{eq:inequality}.

The inequality \eqref{eq:inequality} implies that one can approximate $\langle n_x\rangle_t$ for large $\tau$ as follows,
\eqs{
\braket{n_x}_t 
&\simeq 2\sqrt{\tau}\int_0^{2\gamma \sqrt{\tau}} du \; e^{-u^2}(1-\frac{u^4}{16\gamma^2 \tau}) J_{x-1}(2\sqrt{\tau}u)J_x(2\sqrt{\tau}u)
\\
&\simeq 2\sqrt{\tau}\int_0^{\infty} du \; e^{-u^2}(1-\frac{u^4}{16\gamma^2 \tau}) J_{x-1}(2\sqrt{\tau}u)J_x(2\sqrt{\tau}u),
\label{eq:approximation_n_x}
} 
where the approximation error is bounded by $\mathcal{O}(\tau^{-3/2})$.
In the second line, we extend the upper limit of the integral to infinity, as this modification introduces only an exponentially small error.
To further simplify the expression, we perform the following transformation,
\eqs{
&2\sqrt{\tau}\int_0^{\infty} du \; e^{-u^2} J_{x-1}(2\sqrt{\tau}u)J_x(2\sqrt{\tau}u)
\\
=& 2\sqrt{\tau}\int_0^{\infty} du \; \oint \frac{dz_1}{2\pi i z_1}\oint \frac{dz_2}{2\pi iz_2} e^{-u^2}e^{\sqrt{\tau}u(-z_1+1/z_1-z_2+1/z_2)} 
z_1^{x}z_2^{x-1}
\\
= &2\int_0^{\infty} du \; \oint_{|z_1|\ll 1} \frac{dz_1}{2\pi i z_1}\oint_{|z_2|\gg 1}
 \frac{dz_2}{2\pi iz_2} u e^{-(1+z_1-1/z_2)u^2}e^{\tau(1/z_1-z_2)}z_1^{x}z_2^{x-1}\label{eq:infty_cal_1}
\\
=& \oint_{|z_1|\ll 1} \frac{dz_1}{2\pi i z_1}\oint_{|z_2|\gg 1} \frac{dz_2}{2\pi i}\frac{1}{1+z_1}
\frac{e^{\tau(1/z_1-z_2)}}{z_2-1/(1+z_1)} z_1^x z_2^{x-1}
\\
= &\oint_{|z_1|\ll 1}\frac{dz_1}{2\pi i} \frac{e^{\tau(1/z_1-\frac{1}{1+z_1})}}{1+z_1}\left(\frac{z_1}{1+z_1}\right)^{x-1}\\
= &\oint_{|\eta| < 1}\frac{d\eta}{2\pi i \eta} e^{\tau(\eta+1/\eta-2)}\frac{\eta^x}{1-\eta}.\label{eq:modified_bessel}
}
Here, we use the contour integral expression of the Bessel function in the second line, make the substitutions,
$z_1\to uz_1/\sqrt{\tau}$ and $z_2\to \sqrt{\tau} z_2/u$ in the third line, perform the integration with respect to $u$ by setting
the contours, $|z_1|\ll1$ and $|z_2|\gg 1$ in the forth line, and change the variable $z_1$ to $\eta = z_1/(1+z_1)$ in the last line.
The remaining term in Eq.~\eqref{eq:approximation_n_x} can be also expressed by a similar contour integral.
Thus, we have
\eq{
\langle n_x\rangle_t \simeq & \oint_{|\eta|<1} \frac{d\eta}{2\pi i \eta} \frac{\eta^x}{1-\eta} e^{\tau(\eta+1/\eta-2)}\\
-&\frac{1}{16\gamma^2} \oint \frac{d\eta}{2\pi i \eta} \eta^{x-2}e^{\tau(\eta+1/\eta-2)} [\tau(1-\eta)^3+2\eta(1-\eta)].\label{eq:contour_dw}
}

\begin{figure}[htbp]
    \centering
    \includegraphics[width=0.4\textwidth]{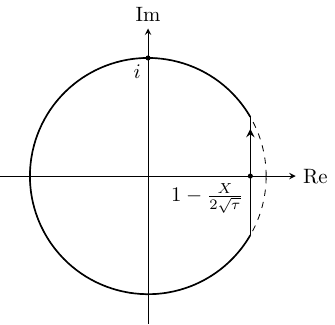} 
    \caption{Schematic illustration of the contour in the asymptotic analysis of Eq.~\eqref{eq:contour_dw}.
    		The arrowed line shows the contour, while the dashed line represents the unit circle for clarity.}
    \label{fig:contour_dw} 
\end{figure}

By performing the saddle point method around $\eta= 1$ in Eq.~\eqref{eq:contour_dw}, we obtain the asymptotic form of 
$\langle n_x\rangle_t$, Eq.~\eqref{eq:asymptotics_dw_app}.
Here, we outline the derivation of the leading term in Eq.~\eqref{eq:asymptotics_dw_app}; the remaining terms can be obtained through similar calculations. 
The leading term arises from the first term in Eq.~\eqref{eq:contour_dw}.
In this term, we deform the contour as depicted in Fig.~\ref{fig:contour_dw}, assuming $X=\mathcal{O}(1)$.
Because the integrand is of order one only when $\eta - 1= \mathcal{O}(1/\sqrt{\tau})$,
we set $\eta = 1-X/(2\sqrt{\tau}) + is$, and perform the saddle point approximation, deriving the leading term,
\begin{align}
\oint_{|\eta|<1}\frac{d\eta}{2\pi i\eta}\frac{\eta^x}{1-\eta} e^{\tau (\eta+1/\eta-2)} &\simeq \int_{-\infty}^{\infty}\frac{ds}{2\pi[X/(2\sqrt{\tau}) - is]}
e^{-\tau s^2} e^{-X^2/4}
\\
&= \int^{\infty}_{-\infty}\frac{ds}{2\pi [X/2-is]} e^{-s^2} e^{-X^2/4}
\\
&= \int^{\infty}_{-\infty} \frac{ds}{\pi}\int_0^{\infty} du\; e^{-s^2} e^{-(X - 2is)u}e^{-X^2/4}
\\
&= \frac{1}{\sqrt{\pi }}\int_{X/2}^{\infty}du \; e^{-u^2}.
\end{align}
Here, we use the identity $1/(X-2is) = \int_0^{\infty} du\;e^{-(X-2is)u}$ for $X\geq 0$ in the third line.
%-------------------------------------------------------------------------------------
%-------------------------------------------------------------------------------------

Next, we consider $\langle Q\rangle_t$ given in Eq.~\eqref{eq:current_dw}.
Similarly to $\langle n_x \rangle$, we can make the following approximation for large $\tau$,
\begin{equation}
\langle Q\rangle_t \simeq 2\tau \int_0^{\infty} du\; e^{-u^2}(1-\frac{1}{16\gamma^2\tau}u^4) u[J_0^2(2\sqrt{\tau}u)+J_1^2(2\sqrt{\tau}u)],
\end{equation}
where the approximation error is bounded by $\mathcal{O}(\tau^{-1})$.
After similar calculations as in the derivation of Eq.~\eqref{eq:modified_bessel},
the above expression can be rewritten in terms of  the modified Bessel function $I_x(\tau)$ as follows,
\begin{equation}
\langle Q\rangle_t \simeq \tau e^{-2\tau}[I_0(2\tau)+I_1(2\tau)] -\frac{1}{8\gamma^2}e^{-2\tau}\left[I_0(2\tau)+\tau(I_1(2\tau)-I_0(2\tau))\right].
\end{equation}
Using the asymptotic form of $I_n(\tau)$\cite{Math_formula}, we have
\begin{equation}
\langle Q\rangle_t \simeq \sqrt{\frac{\tau}{\pi}} -\frac{1}{16\sqrt{\pi \tau}} \left[1+\frac{3}{4\gamma^2}\right].
\end{equation}

Finally, we determine the asymptotic form of $G^{\mathrm{DW}}_{x_1,x_2}(t)$.
In the derivation, we assume that $|x_2-x_1|=\mathcal{O}(1)$ for simplicity.
The exact solution is given in Eq.~\eqref{eq:exact_twopoint_dw}, namely
\begin{equation*}
\begin{split}
G^{\mathrm{DW}}_{x_1,x_2}(t) &=  (-i)^{x_1}i^{x_2}  \int_0^{t} \frac{du}{\sqrt{1-(u/t)^2}} e^{-4\gamma t (1- \sqrt{1-(u/t)^2})} 
\left(\frac{u/t}{\sqrt{1-(u/t)^2}+1}\right)^{|x_2-x_1|}\\
&\times \left[(\sqrt{1-(u/t)^2}+1)J_{x_{Q(1)}}(2u)J_{x_{Q(2)}-1}(2u)+ (\sqrt{1-(u/t)^2}-1)
J_{x_{Q(1)}-1}(2u)J_{x_{Q(2)}}(2u)\right].
\end{split}
\end{equation*}
As in the case of $\langle n_x\rangle_t$, one can make the following approximation for large $\tau$,
\eqs{
G^{\mathrm{DW}}_{x_1,x_2}(t) &\simeq 2 (-i)^{x_1}i^{x_2}\sqrt{\tau} \int_0^\infty du\; e^{-u^2} 
\left(\frac{u}{4\gamma \sqrt{\tau}}\right)^{|x_2-x_1|}J_{x_{Q(1)}}(2\sqrt{\tau}u)J_{x_{Q(2)}-1}(2\sqrt{\tau}u)
\\
&=(-i)^{x_1}i^{x_2}(4\gamma)^{-|x_2-x_1|}\oint \frac{d\eta}{2\pi i \eta} e^{\tau(\eta+1/\eta-2)} \eta^{x_{Q(1)}}(1-\eta)^{|x_{1}-x_{2}|-1}.
}
where the approximation error is bounded by $\mathcal{O}(\tau^{-|x_2-x_1|-1/2})$.
The saddle point method around $\eta=1$ yields
\begin{equation}
G^{\mathrm{DW}}_{x_1,x_2}(t) \simeq \frac{i^{x_2-x_1}}{\sqrt{\pi}} (8\gamma \sqrt{\tau})^{-l} e^{-X_{Q(1)}^2/4}
\sum_{k=0}^{\lfloor (l-1)/2\rfloor} \frac{(-1)^k (l-1)!}{k! (l-1-2k)!}X_{Q(1)}^{l-1-2k}
\label{eq:two_point_dw}
\end{equation}
with the rescaled coordinate $X_{Q(1)}:= x_1/\sqrt{\tau}$ and the relative coordinate $l :=|x_2-x_1|$.
The derivation of Eq.~\eqref{eq:two_point_dw} is similar to that of Eq.~\eqref{eq:asymptotics_dw_app}.

%---------------------------------------------------------------------------------------------------------
\section{Derivation of Eq.~(\ref{eq:exact_twopoint_alt}) }\label{app:derivation_alt}
Here, we derive the exact solution of $G^{\mathrm{Alt}}_{x_1,x_2}(t)$ given in Eq.~\eqref{eq:exact_twopoint_alt}.
The derivation is similar to the case of the domain wall initial condition, however some care is required regarding the sum over the initial condition.

Combining Eq.~\eqref{eq:expansion_alt}, Eq.~\eqref{eq:green}, and Eq.~\eqref{eq:integral}, we have
\begin{equation}
G^{\mathrm{Alt}}_{x_1,x_2}(t)= \sum_{y\in \mathbb{Z}}(-1)^{x_1} \oint_{|z_1|=r}\frac{dz_1}{2\pi i z_1} \oint_{|z_2|=1}\frac{dz_2}{2\pi iz_2} 
e^{-iE(z_1,z_2) t} \frac{s_1-s_2}{s_1-s_2-2\gamma}z_2^{x_{Q(1)}} z_1^{x_{Q(1)}} (z_1z_2)^{-2y}.
\end{equation}
In contrast to the domain wall initial condition, the infinite series cannot be taken directly in the above equation.
Hence, we transform the above equation to the following equation,
\begin{equation}
\begin{split}
G^{\mathrm{Alt}}_{x_1,x_2}(t) = \sum_{y\in \mathbb{Z}} \Big[\delta_{x_1,2y}\delta_{x_2,2y} + i^{x_2-x_1}
\oint \frac{dz_1}{2\pi i z_1} \oint \frac{dz_2}{2\pi iz_2} \int_0^{t}ds \;e^{-4\gamma(t-s)}
\\
\times e^{-(t+s)z_1+(t-s)/z_1}e^{-(t-s)z_2+(t+s)/z_2}
(z_1z_2)^{-2y} z_2^{x_{Q(1)}} z_1^{x_{Q(2)}} 2(c_1-c_2)\Big].\label{eq:derrida_trick_alt}
\end{split}
\end{equation}
The derivation is similar to that of Eq.~\eqref{eq:derrida_trick}.
Since the radius of the contour can be chosen arbitrary in Eq.~\eqref{eq:derrida_trick_alt},
we choose $|z_1|< 1,\; |z_2|=1$ for $y\leq 0$ and $|z_1|>1,\;|z_2|=1$ for $y \geq 1$.
Then, one can take the infinite series and obtain
\begin{equation*}
\begin{split}
&G^{\mathrm{Alt}}_{x_1,x_2}(t) = \frac{1+(-1)^{x_1}}{2}\delta_{x_1,x_2} 
\\
&+i^{x_2-x_1}\oint_{|z_1|<1} \frac{dz_1}{2\pi i z_1} \oint_{|z_2|=1} \frac{dz_2}{2\pi iz_2} \int_0^{t}ds \;e^{-4\gamma(t-s)}
\frac{e^{-(t-s)(z_2-1/z_1)+(t+s)(1/z_2-z_1)}}{1+z_1z_2}\frac{z_2-z_1}{z_1z_2} z_2^{x_{Q(1)}}z_1^{x_{Q(2)}}
\\
&-i^{x_2-x_1}\oint_{|z_1|>1} \frac{dz_1}{2\pi i z_1} \oint_{|z_2|=1} \frac{dz_2}{2\pi iz_2} \int_0^{t}ds \;e^{-4\gamma(t-s)}
\frac{e^{-(t-s)(z_2-1/z_1)+(t+s)(1/z_2-z_1)}}{1+z_1z_2}\frac{z_2-z_1}{z_1z_2} z_2^{x_{Q(1)}}z_1^{x_{Q(2)}}.
\end{split}
\end{equation*}
Note that the integrands of the second and third terms in the above equation are the same.
By shrinking the $z_1$-contour in the third term, the second and third terms cancel each other,
leaving only the contribution of the pole $z_1 = -1/z_2$.
Thus, we have
\begin{equation}
\begin{split}
G^{\mathrm{Alt}}_{x_1,x_2}(t)= &\frac{1+(-1)^{x_1}}{2}\delta_{x_1,x_2}
-(-1)^{x_{Q(2)}} i^{x_2-x_1}\oint\frac{dz}{2\pi iz}
\\
&\times \int_0^t ds \;e^{-4\gamma(t-s)} e^{-2(t-s)z+2(t+s)/z}(z+1/z) z^{x_{Q(1)}-x_{Q(2)}}.
\end{split}\label{eq:start_asymptotics}
\end{equation}
After similar calculations to that for the domain wall initial condition, the above equation can be rewritten in terms of 
the Bessel function $J_{\nu}(u)$ as follows,
\begin{equation}
\begin{split}
G^{\mathrm{Alt}}_{x_1,x_2}(t)&=\frac{1+(-1)^{x_1}}{2}\delta_{x_1,x_2}
 + (-1)^{x_{Q(1)}} i^{x_2-x_1}
\int_0^{t}\frac{du}{\alpha(u)}e^{-4\gamma t(1-\alpha(u))}
\\ &\times
[(\alpha(u)+1)J_{|x_2-x_1|-1}(4u) -(\alpha(u)-1)J_{|x_2-x_1|+1}(4u)]\left(\frac{u/t}{1+\alpha(u)}\right)^{|x_2-x_1|}
\end{split}
\end{equation}
with $\alpha(u)= \sqrt{1-(u/t)^2}$.

%------------------------------------------------------------------------------------------------------------------------
\section{Asymptotic analysis for the alternating initial condition} \label{app:asymptotics_alt}
In this section, we determine the asymptotic form of $G^{\mathrm{Alt}}_{x_1,x_2}(t)$ for large $t$.
We show that the relaxation of $G^{\mathrm{Alt}}_{x_1,x_2}(t)$ undergoes a dynamical transition from oscillatory decay to
over-damped decay as $\gamma $ increases.

As the start point of calculations, instead of Eq.~\eqref{eq:exact_twopoint_alt}, we use Eq.~\eqref{eq:start_asymptotics}.
In the following, we analyze the asymptotic form of $I$ which is defined as 
\begin{equation}
I := \oint\frac{dz}{2\pi iz} \int_0^t ds \;e^{-4\gamma(t-s)} e^{-2(t-s)z+2(t+s)/z}(z+1/z) z^{x_{Q(1)}-x_{Q(2)}}.
\end{equation}
We set the contour with $|z|\gg 1$ in the above equation and integrate with respect to $s$.
Then, we have
\begin{equation}
\begin{split}
I = \frac{1}{2}\oint_{|z|\gg1}\frac{dz}{2\pi iz} \frac{c(z) }{c(z)+\gamma} e^{4t/z} z^{x_{Q(1)}-x_{Q(2)}}
-\frac{1}{2}\oint_{|z|\gg1}\frac{dz}{2\pi i z} \frac{c(z)}{c(z)+\gamma}e^{2t(-z+1/z)-4\gamma t}z^{x_{Q(1)}-x_{Q(2)}}\label{eq:app_alt_2}
\end{split}
\end{equation}
with $c(z):= (z+1/z)/2$.
After the substitution $z_1\to 1/z_1$, the first term in the above equation becomes 
\begin{equation}
\text{the first term} = \frac{1}{2}\oint_{|z|\ll 1}\frac{dz}{2\pi iz} \frac{c(z) }{c(z)+\gamma} e^{4t z} z^{x_{Q(2)}-x_{Q(1)}}.\label{eq:app_alt_3}
\end{equation}
Since the poles of $1/(c(z)+\gamma)$ are outside the contour, one can conclude that the first term is $\delta_{x_1,x_2}/2$ 
from the residue theorem.
 
%---------------------------------------------------------------------------------------------------------------
\begin{figure}[htbp]
    \centering
    \begin{subfigure}{0.32\textwidth}
        \centering
        \includegraphics[width=\textwidth]{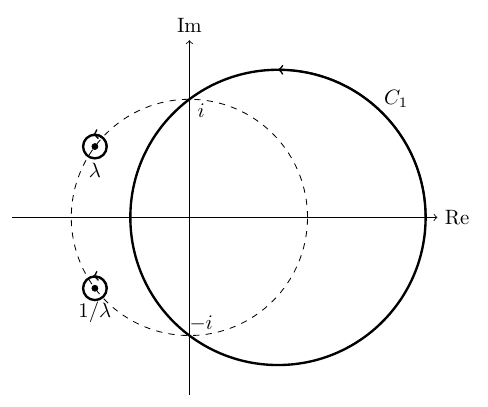}
        \caption{$\gamma < 1$}
        \label{subfig:subfig1}
    \end{subfigure}
    \hfill
    \begin{subfigure}{0.32\textwidth}
        \centering
        \includegraphics[width=\textwidth]{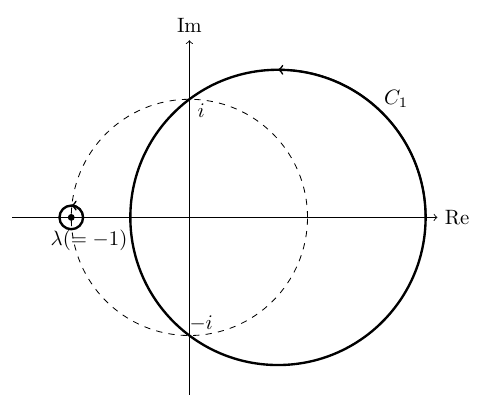}
        \caption{$\gamma$ =1}
        \label{subfig:subfig2}
    \end{subfigure}
    \hfill
    \begin{subfigure}{0.32\textwidth}
        \centering
        \includegraphics[width=\textwidth]{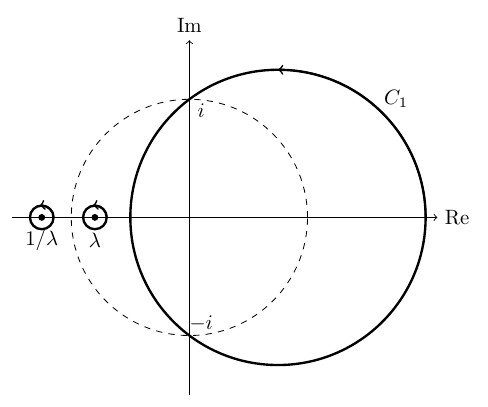}
        \caption{$\gamma>1$}
        \label{subfig:subfig3}
    \end{subfigure}
    \caption{Schematic illustrations of the contours.
    The arrowed lines represent the contours.
    The dots represent the poles $\lambda,\;1/\lambda$ of $\frac{1}{c(z)+\gamma}$. 
    Note that the poles depend on $\gamma$ and become 
    a second-order pole for $\gamma=1$.
    The dashed lines represent a unit circle for clarity.}
    \label{fig:contour}
\end{figure}
%---------------------------------------------------------------------------------------------------------------

To evaluate the second term, we deform the contour as depicted in Fig.~\ref{fig:contour}, 
separating the contributions from the poles of $1/(c(z)+\gamma)$ and from the contour $C_1$.
Note that the poles of $1/(c(z)+\gamma)$ is given by $\lambda$ and $1/\lambda$ 
with $\lambda := -\gamma + \sqrt{\gamma^2-1}$,
and the poles become a second-order pole when $\gamma =1$.
We first evaluate the contour integral on $C_1$.
Since $\mathrm{Re}[-z+1/z] < 0$ on $C_1$ except for the case $z= \pm i$,
the contour integral on $C_1$ can be evaluated using the saddle point method around $z= \pm i$.
Its leading term is given by $\mathcal{O}(1/\sqrt{t})e^{-4\gamma t}$.
On the other hand,
the pole contributions can be exactly calculated by the residue theorem.
Comparing these results,
one finds that the pole contributions provide the leading term.
Eventually, we have
\begin{equation}
I -\frac{\delta_{x_1,x_2}}{2} \simeq 
\begin{cases}
 \frac{-\gamma}{2\sqrt{\gamma^2-1}} \left( \frac{-1}{\gamma+\sqrt{\gamma^2-1}}\right)^{|x_1-x_2|} 
 e^{-4\gamma t(1-\sqrt{1-1/\gamma^2})}
 &\text{for} \;\; \gamma >1,
 \\
-4t(-1)^{|x_1-x_2|}e^{-4\gamma t} &\text{for}\;\; \gamma =1,
\\
\frac{-\gamma}{\sqrt{1-\gamma^2}}(-1)^{|x_1-x_2|} 
\sin \left[4t\sqrt{1-\gamma^2} - |x_1-x_2|\cos^{-1}(\gamma)\right] e^{-4\gamma t}
&\text{for} \;\; 0< \gamma <1.
\end{cases}
\end{equation}
Substituting the above equation into Eq.~\eqref{eq:start_asymptotics} yields the asymptotic form of 
$G^{\mathrm{Alt}}_{x_1,x_2}(t)$, 
\begin{equation}
G^{\mathrm{Alt}}_{x_1,x_2}(t)\simeq 
\begin{cases}
\frac{1}{2}\delta_{x_1,x_2} 
+
\frac{\gamma (-1)^{x_1} }{2\sqrt{\gamma^2-1}}
\left( \frac{i}{\gamma+\sqrt{\gamma^2-1}}\right)^{|x_1-x_2|} 
 e^{-4\gamma t(1-\sqrt{1-1/\gamma^2})}  &\text{for} \;\; \gamma >1,
\\
\frac{1}{2}\delta_{x_1,x_2} + 4 (-1)^{x_1} i^{|x_1-x_2|} t e^{-4\gamma t} &\text{for}\;\; \gamma =1,
\\
\frac{1}{2}\delta_{x_1,x_2} +
\frac{\gamma(-1)^{x_1} }{\sqrt{1-\gamma^2}} 
i^{|x_1-x_2|}
\sin\left[ 4t\sqrt{1-\gamma^2}-|x_1-x_2|\cos^{-1}(\gamma) \right] e^{-4\gamma t} &\text{for} \;\; 0< \gamma <1.
\end{cases}
\end{equation}

%------------------------------------------------------------------------------------------------------------
\section{Strong dephasing limit of the average density}\label{app:zeno}
We consider the large $\gamma$ limit of the average density under the domain wall and the alternating initial conditions.

\subsection{Domain wall initial condition}
We derive Eq.~\eqref{eq:zeno_dw} for $\gamma \gg1$ (with finite $\tau$).
From Eq.~\eqref{eq:inequality}, one can make the following approximation,
\eqs{
\braket{n_x}_t 
&\simeq 2\sqrt{\tau} \int_0^{2\gamma \sqrt{\tau}} du\;e^{-u^2} (1-\frac{u^4}{16\gamma^2\tau}) J_{x-1}(2\sqrt{\tau}u)J_x(2\sqrt{\tau}u)
\\
&\simeq 2\sqrt{\tau}\int_0^{\infty} du\;e^{-u^2} (1-\frac{u^4}{16\gamma^2\tau}) J_{x-1}(2\sqrt{\tau}u)J_x(2\sqrt{\tau}u)
\\
&= \oint_{|\eta|<1} \frac{d\eta}{2\pi i \eta} \frac{\eta^x}{1-\eta} e^{\tau(\eta+1/\eta-2)}
-\frac{1}{16\gamma^2} \oint \frac{d\eta}{2\pi i \eta} \eta^{x-2}e^{\tau(\eta+1/\eta-2)} [\tau(1-\eta)^3+2\eta(1-\eta)],
}
where the approximation error is bounded by $\order{\gamma^{-4}}$. 
In the second line, we extend the upper limit of the integral to infinity, as this modification introduces only an exponentially small error,
and in the third line, we perform the exactly same calculation in the derivation of Eq.~\eqref{eq:contour_dw}.

\subsection{Alternating initial condition}
We derive Eq.~\eqref{eq:zeno_alt} for $\gamma \gg 1$ (with finite $\tau$).
From Eqs.~\eqref{eq:start_asymptotics}, \eqref{eq:app_alt_2}, and \eqref{eq:app_alt_3}, we have
\eqs{
\braket{n_x}_t &= \frac{1}{2} + \frac{(-1)^x}{2} \oint_{|z|\gg 1} \frac{dz}{2\pi i z}\frac{c(z)}{c(z)+\gamma}e^{4\gamma \tau(-z+1/z)-8\gamma^2\tau}
\\
&=\frac{1}{2}  + \frac{(-1)^x}{2} \frac{\gamma \; e^{-8\gamma^2\tau(1- \sqrt{1-1/\gamma^2})}}{\sqrt{\gamma^2-1}} 
+ \frac{(-1)^x}{2} \oint_{|z|= 1} \frac{dz}{2\pi i z}\frac{c(z)}{c(z)+\gamma}e^{4\gamma \tau(-z+1/z)-8\gamma^2\tau}
\\
&\simeq \frac{1}{2}  + \frac{(-1)^x}{2} \frac{\gamma \; e^{-8\gamma^2\tau(1- \sqrt{1-1/\gamma^2})}}{\sqrt{\gamma^2-1}} 
\\
&= \frac{1}{2} + \frac{(-1)^x}{2} e^{-4\tau} \left[ 1 +\frac{1-2\tau}{2\gamma^2}\right] +\order{\gamma^{-4}}.
}
Here, we apply the residue theorem in the second line, and omit the term that is exponentially small with respect to $\gamma\gg1$ in the third line. 

\bibliographystyle{unsrt}
\bibliography{ref}

\end{document}